\documentstyle{article}

\newtheorem{htwierdzenie}{Theorem}[section]
\newtheorem{hfakt}[htwierdzenie]{Proposition}
\newtheorem{hlemat}[htwierdzenie]{Lemma}
\newtheorem{hwniosek}[htwierdzenie]{Corollary}
\newtheorem{hdefinicja}[htwierdzenie]{Definition}
\newtheorem{hUwaga}[htwierdzenie]{Remark}
\newcommand{\hhtw}{{\bf\hspace{-0.45em}.\@\hspace{0.45em}}}
\newenvironment{twierdzenie}{\begin{htwierdzenie} \hhtw}{\end{htwierdzenie}}
\newenvironment{fakt}{\begin{hfakt} \hhtw}{\end{hfakt}}
\newenvironment{lemat}{\begin{hlemat} \hhtw}{\end{hlemat}}
\newenvironment{wniosek}{\begin{hwniosek} \hhtw}{\end{hwniosek}}
\newenvironment{definicja}{\begin{hdefinicja} \hhtw}{\end{hdefinicja}}
\newenvironment{Uwaga}{\begin{hUwaga} \hhtw}{\end{hUwaga}}

\newcommand{\dow}{
\par
{\em Proof.}\hspace{2em}}

\newcommand{\refdow}[1]{
\par
{\em Proof #1}\hspace{2em}}

\newcommand{\cnd}{
   \nopagebreak \mbox{\rule{1cm}{0cm}}
   \nopagebreak \hspace*{\fill} \nolinebreak
   \nopagebreak \raisebox{-1.5ex}{\large Q.\ E.\ D.\@}
\par
\vspace{3ex}
}

\newcommand{\dowspace}{
\par
\vspace{1.5ex}
}

\newcommand{\C}{\mbox{{\bf C}}{}}

\newcommand{\N}{\mbox{{\bf N}}{}}
\newcommand{\Z}{\mbox{{\bf Z}}{}}
\setbox3=\hbox{$\perp$}
\newcommand{\ti}{{\,\smash{\bigcirc}\hskip-9.03pt\raise0.05pt\copy3\,}}
\setbox1=\hbox{$\top$}
\newcommand{\tp}{\rule{0em}{1.5ex}%
{\,\smash{\bigcirc}\hskip-9pt\raise-1.7pt\copy1\,}%
\rule{0.15em}{0ex}}
\newcommand{\tptil}{\rule{0em}{1.5ex}%
{\,\smash{\bigcirc}\hskip-8.93pt\raise-1.7pt\copy1\,}%
\rule{0.15em}{0ex}}
\newcommand{\tpt}{\widetilde{\tptil}}
\setbox2=\hbox{\scriptsize$\top$}
\newcommand{\tps}{{\,\smash{\bigcirc}\hskip-7.09pt\raise-1.4pt\copy2\,}}
\newcommand{\tpstil}{{\,\smash{\bigcirc}\hskip-7.17pt\raise-1.4pt\copy2\,}}

\newcommand{\Subrep}{{\raise0.8pt\hbox{$\scriptstyle\subset$}}}
\newcommand{\subrep}{{\raise1.3pt\hbox{$\scriptscriptstyle\subset$}}}

\newcommand{\Suprep}{{\raise0.8pt\hbox{$\scriptstyle\supset$}}}
\newcommand{\suprep}{{\raise1.3pt\hbox{$\scriptscriptstyle\supset$}}}
\newcommand{\image}{\mbox{\rm Im}\,}
\newcommand{\spn}{\mbox{\rm span}}
\newcommand{\In}{{\raise1pt\hbox{$\scriptscriptstyle\in$}}}
\newcommand{\nI}{{\raise1pt\hbox{$\scriptscriptstyle\ni$}}}
\newcommand{\la}{\lambda}
\newcommand{\al}{\alpha}
\newcommand{\be}{\beta}
\newcommand{\ga}{\gamma}
\newcommand{\de}{\delta}
\newcommand{\si}{\sigma}
\newcommand{\ot}{\otimes}

\newcommand{\restr}[2]{#1%
\mbox{\raisebox{-1ex}{$\mid$}\raisebox{-1.5ex}{$\scriptstyle #2$}}}
\newcommand{\quot}[2]{%
\mbox{\raisebox{.5ex}{$#1$}\hspace{-0.1em}$\slash$%
\hspace{-0.15em}\raisebox{-.5ex}{$#2$}}}
\newcommand{\tr}{\mbox{\rm Tr}\,}

\newcommand{\hlabel}[1]{\label{#1}}

\newcommand{\hsection}[1]{
\setcounter{equation}{0}
\section[\hspace{-0.9em}.\@\hspace{0.9em}#1]%
{\hspace{-1.05em}.\@\hspace{1.05em}#1}
}

\newcommand{\hsubsection}[1]{
\subsection[\hspace{-1.0em}.\@\hspace{1.0em}#1]%
{\hspace{-1.05em}.\@\hspace{1.05em}#1}
}

\newcommand{\hasection}[1]{
\setcounter{equation}{0}
\section[\hspace{-0.6em}.\@\hspace{0.6em}#1]%
{\hspace{-1.05em}.\@\hspace{1.05em}#1}
}

\newcommand{\jnd}{\frac{1}{2}}
\newcommand{\fact}{\mbox{\rm Fact}}
\newcommand{\id}{\mbox{\rm id}}

\newcommand{\iInv}[1]{\mbox{\footnotesize\hspace{0.4em}$m(#1)$}}
\newcommand{\iINV}[1]{\mbox{\hspace{0.4em}$m(#1)$}}

\newcommand{\WT}{\widetilde{\rule{1.5em}{0ex}}}

\newcommand{\uwaga}{{\em Remark.}\hspace{1em}}

\newcommand{\tmp}{}

\newcommand{\psiP}{\psi_{\hspace{-0.5mm}\raisebox{-1mm}{$\scriptstyle P$}}}
\newcommand{\Kom}{\triangle}
\newcommand{\SetPow}[1]{\# #1 }
\newcommand{\Haar}{\mbox{h}}
\newcommand{\ABHom}{\mbox{F}}
\newcommand{\flip}{\tau}
\newcommand{\Uqsl}{U_qsl(2)}
\newcommand{\pair}[2]{\mbox{$ \langle #1 \:, #2 \rangle $}}

\newcommand{\Jm}{J^-}
\newcommand{\Jp}{J^+}

\newcommand{\Jpm}{J^{\pm}}
\newcommand{\Kp}{q^{\jnd H}}
\newcommand{\Km}{{ q^{-\jnd H} }}
\newcommand{\Kpm}{{ q^{\pm\jnd H} }}

\newcommand{\UL}{U_L}
\newcommand{\UA}{U_{\AP}}

\newcommand{\AP}{{\cal A}}

\begin{document}


\hfill hep-th/9405079

\vspace*{1cm}
\begin{center}
  \Huge On representation theory of quantum $SL_q(2)$ groups
  at roots of unity
\end{center}

\vspace{0.5cm}
\begin{center}
{\Large P. Kondratowicz and P. Podle\'{s}}
 \\[0.5cm]
  Department of Mathematical Methods in Physics, \\
  Faculty of Physics,
  University of Warsaw, \\
  Ho\.{z}a 74, 00-682 Warsaw, Poland, \\
  \begin{tabular}{ccc}
    e-mail: & {\sf kondrat@fuw.edu.pl } & {\sf podles@fuw.edu.pl }
  \end{tabular}
\end{center}

\vspace{0.5cm}

\begin{abstract}
 \small
 Irreducible representations of quantum groups $SL_q(2)$ (in
 Woronowicz' approach) were classified in J.Wang, B.Parshall, Memoirs AMS
 439 in the~case of $q$ being an~odd root of unity.
 Here we find the~irreducible representations for all roots of unity
 (also of an~even degree), as
 well as describe "the~diagonal part" of tensor product of
 any two irreducible representations.
 An~example of not completely reducible representation is given.
 Non--existence of Haar functional is proved.
 The~corresponding representations of universal enveloping
 algebras of Jimbo and Lusztig are provided.
 We also recall the~case of general~$q$.
 Our computations are done in explicit way.
\end{abstract}

\vspace{0.5cm}

\newpage

\setcounter{section}{-1}
\hsection{Introduction}
\hlabel{i:introduction}

The~quantum $SL(2)$ group is by definition a~quantum group $(A, \Kom)$
that has the~same
representation theory as $SL(2)$, i.e. all nonequivalent irreducible
representations are $u^s$, $s=0,\jnd, 1, \ldots$ \ such that
\renewcommand{\arraystretch}{0.8}%
\[
  \dim u^s = 2s+1 \:, \;\;\;\;\;\;\;
  u^t \tp u^s  \approx
  \raisebox{-1.2ex}{$
    \begin{array}{c}
      \scriptstyle {t+s} \\
      \oplus \\
      \scriptstyle {r=|t-s|} \\
      \mbox{\scriptsize step $ = 1 $}
    \end{array}
  $}
  u^r \:,
\]
\renewcommand{\arraystretch}{1}%
$s,t=0,\jnd, 1, \ldots$ and matrix elements of
$u^{\jnd}$ generate $A$ as an algebra with unity~$I$.

Putting $t,s=\jnd$ in the~above formula one can see that
there must exist nonzero intertwiners
$E \in \mbox{\rm Mor}(u^0,u^{\jnd} \tp u^{\jnd})$
and $E' \in \mbox{\rm Mor}(u^{\jnd} \tp u^{\jnd},u^0)$.
The~operator $E$ may be identified with a~tensor $E \in K \ot K$,
whereas the~operator $E'$ with a~tensor $E' \in K^{\ast} \ot K^{\ast}$,
where $K\approx\C^2$ is the~carrier vector space of $u^{\jnd}$.

The~classification of quantum $SL(2)$ groups (described in the~introduction of
\cite{b:SL2} and repeated here)
is based on consideration of these tensors. There are three cases:
\begin{enumerate}
  \item The~rank of the~symmetric part of $E$ is~$0$. Then
    \begin{eqnarray*}
       & E = e_1 \ot e_2 - e_2 \ot e_1 \:, \\
       & E' = e^1 \ot e^2 - e^2 \ot e^1 \:,
    \end{eqnarray*}
    where $e_1$, $e_2$ is a basis in $K$, while $e^1$, $e^2$
    is the~dual basis in $K^{\ast}$.
    This case corresponds to the~undeformed (classical) $SL(2)$.
  \item The~rank of the~symmetric part of $E$ is~$1$. Then there exists
    a~basis $e_1$, $e_2$ in $K$ such that
    \begin{eqnarray*}
      & E = e_1 \ot e_2 - e_2 \ot e_1 + e_1 \ot e_1 \:, \\
      & E' = - e^1 \ot e^2 + e^2 \ot e^1 + e^2 \ot e^2 \:.
    \end{eqnarray*}
    This case has been considered e.g. in \cite{b:SL2}.
  \item The~rank of the~symmetric part $E$ is~$2$. Then there exists
    \\
    $q \in \C \setminus \{ 0,1,\mbox{roots of unity}\} \cup \{-1\}$
    and a~basis $e_1$, $e_2$ in $K$ such that
    \begin{eqnarray*}
      & E = e_1 \ot e_2 - q \, e_2 \ot e_1 \;, \\
      & E' = e^1 \ot e^2 - q \, e^2 \ot e^1 \;.
    \end{eqnarray*}
    This~case has been considered in~\cite{b:wykladWor} and is recalled in
    section~\ref{c:clasicalcase}. \\
    For $q=1$ one obtains the~case~1.
\end{enumerate}
 In all these cases $A$ is the~algebra with unity $I$
 generated by matrix elements of $u=u^{\jnd}$ satisfying the~relations
\[
 (u\tp u)E=E  \:, \;\;\;\; E'(u\tp u)=E' \:.
\]

When $q$ is a root of unity and $q\neq \pm 1$,
the~representation theory of the~case~3.\
is essentially different from that of $SL(2)$.
In~this case these objects are ambiguously called quantum $SL_q(2)$
groups at roots of unity.
They are considered in the~present paper in section~\ref{m:maincase}.

The basic facts concerning quantum qroups and their representations are
recalled in the~appendix~\ref{d:basicqg}. The~quotient representations
and the~operation~$\WT$ are investigated in the~appendix~\ref{q:qkor}.

\subsubsection*{Basic notions}

The~degree of a~root of unity $q\in\C$ is the~least natural number~$N$
such that $q^N=1$. In~the~following we assume $N\geq 3$, i.e.
$q=\pm 1$ are not roots of unity in our sense. We put
\[
  \begin{array}{c}
    N_0 = \left\{
    \begin{array}{c@{\hspace{1cm}}l}
      N  &  \mbox{if $N$ is odd,} \\
      \frac{N}{2} &  \mbox{if $N$ is even,} \\
      + \infty    &  \mbox{if $q$ is not a~root of unity.}
    \end{array}
    \right.
  \end{array}
\]
We denote by $\N$ the~set of natural numbers $\{0,1,2,\ldots\}$.

\subsubsection*{Results}

 When $q$ is a~root of unity, then all nonequivalent
 irreducible representations of quantum $SL_q(2)$ group are
 $v^t \tp u^s$, \  $t=0,\jnd,1,\ldots$, \
 $s=0,\jnd,1,\ldots,\frac{N_0}{2}-\jnd$.
 Here
\[
  u = u^{\jnd} =
     \left(
        \matrix{ \al & \be \cr \ga & \de \cr }
     \right)
  \;\;\; \mbox{ and }  \;\;\;
  v = v^{\jnd} =
     \left(
        \matrix{ \al^{N_0} & \be^{N_0} \cr \ga^{N_0} & \de^{N_0} \cr }
     \right) \:.
\]
 Moreover, $v$~is a fundamental representation
 of a~quantum group isomorphic to $SL_{q'}(2)$, where $q'=q^{N_0^2}=\pm 1$.
 The~following formulae hold
 \[
 \begin{array}{c}
   v^t \tp v \approx v^{t-\jnd} \oplus v^{t+\jnd} \:,\;\;\;
   \dim v^t = 2 t + 1 \:,
 \\
   u^s \tp u \approx u^{s-\jnd} \oplus u^{s+\jnd} \:,\;\;\;
   \dim u^s = 2 s + 1 \:.
 \end{array}
 \]
 $t=0,\jnd,1, \ldots$, $s=\jnd,1,\ldots,\frac{N_0}{2}-1$.
 Let us summarize the~main results of the~paper. The~basic decomposition
 of the~representation $u^{\frac{N_0}{2}-\jnd} \tp u$ is described by
 \renewcommand{\arraystretch}{1.4}%
 \[
  u^{\frac{N_0}{2}-\jnd} \tp u \approx
  \left(
  \begin{array}{c|c|c}
     u^{\frac{N_0}{2}-1} & \ast & \ast                \\ \hline
     0                   & v    & \ast                \\ \hline
     0                   & 0    & u^{\frac{N_0}{2}-1}
  \end{array}
  \right) \:.
 \]
 \renewcommand{\arraystretch}{1}%
Moreover, the~elements denoted by three stars and the~matrix elements of
 the~representations $u^{\frac{N_0}{2}-1}$ and~$v$ are
 linearly independent.
 Thus the~representation $u^{\frac{N_0}{2}-\jnd} \tp u$
 is not completely reducible.

 One has
\[
  v^t \tp u^s \approx u^s \tp v^t \:.
\]
We also describe the~"diagonal part" of tensor product of~any two
irreducible representations of~$SL_q(2)$.

In sections~\ref{e:EnvelopingAlgebra}---\ref{l:EnvelopingAlgAcor2Lusz}
we describe representations of~universal enveloping algebras
of~Jimbo and Lusztig corresponding to
the~irreducible representations of~$SL_q(2)$.

\uwaga
Let $N$ be odd. Then the~classification of~irreducible representations
of~$SL_q(2)$ is given in~\cite{b:Parshall}. Description of~$v$ as
a~fundamental representation of $SL_1(2)$ is also contained
in~\cite{b:Parshall}.
In the~present paper we consider also $N$ even, prove our results in
an~explicit way and also show other results concerning representation theory
of~$SL_q(2)$ (see sections~\ref{m:BasicDecomposition}.,
\ref{o:MoreAboutIrrRep}.---\ref{l:EnvelopingAlgAcor2Lusz}.).


\hsection{The~general case}
\hlabel{c:clasicalcase}

In this section we recall the~theory of quantum $SL_q(2)$ groups
for general~$q \in \C \setminus \{0\}$, see~\cite{b:wykladWor},
\cite{b:SU2} and \cite{b:Parshall}.

In the~present section $q \in \C \setminus \{0\}$
unless it is said otherwise.

We set $K=\C^2$ with canonical basis~$e_1, \: e_2$.
We fix linear mappings $E : \C \rightarrow K \ot K$ and
$E' : K \ot K \rightarrow \C$ in the~same way as in the~case~3.\ of
the~classification of quantum $SL(2)$ groups in section~\ref{i:introduction}.
\begin{eqnarray}
\hlabel{c:defE}
  &  E(1) = e_1 \ot e_2 - q e_2 \ot e_1 \: ,  \\
\hlabel{c:defE'}
  &
  \renewcommand{\arraystretch}{1.3}
  \begin{array}{cc}
      E'( e_1 \ot e_1 ) = 0 \:, &  E'( e_2 \ot e_2 ) = 0 \:, \\
      E'( e_1 \ot e_2 ) = 1 \:, &  E'( e_2 \ot e_1 ) = -q \:.
  \end{array}
  \renewcommand{\arraystretch}{1}
\end{eqnarray}

\begin{definicja}
\hlabel{c:EE'def}
  $A$ is the~universal algebra with unity generated
  by $ u_{i j}$, \ $i,j = 1,2 $ satisfying
  \begin{equation}
    \hlabel{c:EE'equ}
   (u\tp u)E=E  \;\;\; \mbox{and} \;\;\; E'(u\tp u)=E' \:,
  \end{equation}
  where $u=(u_{ij})_{i,j=1}^2$.
\end{definicja}

Setting
\begin{equation}
  \hlabel{c:udef}
  u^{\jnd} = u =
     \left( \matrix{ \al & \be \cr \ga & \de \cr } \right)
     \in M_{2 \times 2}(A) \:,
\end{equation}
the~relations~(\ref{c:EE'equ}) take the~form
\renewcommand{\arraystretch}{1.3}%
\begin{equation}
  \hlabel{c:relations}
  \begin{array}{c}
    \al \be = q \be \al \:,\;\; \al \ga = q \ga \al \:, \\
    \be \de = q \de \be \:,\;\; \ga \de = q \de \ga \:, \\
    \be \ga = \ga \be  \:, \\
    \al \de - q \be \ga = I \:,\;\; \de \al - \frac{1}{q} \be \ga = I \:.
  \end{array}
\end{equation}
\renewcommand{\arraystretch}{1}%

Using~(\ref{c:EE'equ}) one can easily prove the~following
\begin{fakt}
  There exists unique structure of Hopf algebra in $A$ such that
  $u$ is a~representation (this~representation is called fundamental).
\end{fakt}

Let
\begin{eqnarray}
&
  \hlabel{c:alk}
  \al_k = \left\{
    \begin{array}{ll}
      \al^k    & \mbox{for $k \geq 0$} \\
      \de^{-k} & \mbox{for $k < 0$} \;.
    \end{array}
  \right.
& \\
&
 \hlabel{c:Akdef}
 A_k = \spn \left\{ \al_s \be^m \ga^n \; : \;
   s \in \Z,\; m,n \in \N, \; |s| + m + n \leq k \right\} \:,
   \;\;\; k \in \N \:.
&
\end{eqnarray}

\begin{fakt}
  \hlabel{c:baza}
  Elements of the~form
  \begin{equation}
    \hlabel{c:elembazy}
    \al_k \ga^m \be^n \:, \;\;\;  k \in \Z \:, \;\; m,n \in \N \:,
  \end{equation}
  form a~basis of the~algebra $A$.
  Moreover
  \begin{equation}
    \hlabel{c:Akdim}
    \dim A_k = \sum_{l=0}^k ( l + 1 )^2.
  \end{equation}
\end{fakt}

The~proof is given in~\cite{b:Lorentz} (Proposition~4.2.).

Elements of the~form
\begin{equation}
  \hlabel{c:elementybazy2}
  \al^k \de^l \be_m \;, \hspace{2em}  k, l \in \N \:,\;\;\; m \in \Z
\end{equation}
are also a~basis of the~algebra $A$, where by definition
\begin{equation}
  \hlabel{c:bem}
  \be_m = \left\{
    \begin{array}{ll}
      \be^m    & \mbox{for $m \geq 0$} \\
      \ga^{-m} & \mbox{for $m < 0$} \;.
    \end{array}
  \right.
\end{equation}
This follows from the~fact that
each of the~elements~(\ref{c:elementybazy2}) of a~given degree
(a~degree of an element (\ref{c:elembazy}) or (\ref{c:elementybazy2})
is the~sum of absolute values of its indices)
can be expressed as a~finite linear combination
of~the~elements~(\ref{c:elembazy}) of
the~same or less degree and from the~fact that
the~numbers of both kinds of elements of the~same degree are equal.

The~above consideration shows that
\[
 A_p = \spn \left\{ \al^k \de^l \be_m \; : \;
       k + l + |m| \leq p, \;\; k, l \in \N, \;\; m \in \Z
            \right\} \;\;\; \mbox{for \ $p \in \N$} \:.
\]

In virtue of~(\ref{c:EE'equ}) $ E E', \id \in Mor(u^{\tps 2}, u^{\tps 2}) $.
One has $(\id-E E')(e_i \ot e_j)=e_j \ot e_i$, $i,j=1,2$ for $q=1$.
It means that $\id-E E'$ is equivalent to a~transposition in this case.
Using this intertwiner one can investigate symmetric and antisymmetric
vectors (vectors such that $(\id-E E')v=\pm v$~).
For general~$q$ we are interested in intertwiners of the~form
$\si = \id+\la E E'$, satisfying the~condition
\begin{equation}
  \hlabel{c:si1}
  (\si \ot \id ) (\id \ot \si ) (\si \ot \id ) =
  (\id \ot \si ) (\si \ot \id ) (\id \ot \si )
\end{equation}
After some calculations one gets
$\la = -q^{-2}, -1$.
Taking $\la = -1$ one obtains
\begin{displaymath}
  \si = \id - E E'
\end{displaymath}
(the~other value of $\la$ corresponds to $\si^{-1}$).
Using (\ref{c:defE}) and (\ref{c:defE'}) one has
\renewcommand{\arraystretch}{1.3}%
\begin{equation}
  \hlabel{c:si()}
  \begin{array}{l}
    \si( e_1 \ot e_1 ) = e_1 \ot e_1  \:, \\
    \si( e_1 \ot e_2 ) = q \, e_2 \ot e_1  \:, \\
    \si( e_2 \ot e_1 ) = q \, e_1 \ot e_2 + (1-q^2) \, e_2 \ot e_1 \:, \\
    \si( e_2 \ot e_2 ) = e_2 \ot e_2  \:.
  \end{array}
\end{equation}
\renewcommand{\arraystretch}{1}%
It can be easily found that
\begin{equation}
  \hlabel{c:si2}
  \si^2 = (1-q^2) \si + q^2 \:,
\end{equation}
i.e. $(\si - \id)(\si+q^2) = 0$.

The~eigenvalue~$1$ corresponds to symmetric vectors
\[
  e_1 \ot e_1 \:,\;\;\;\; q \, e_1 \ot e_2 + e_2 \ot e_1 \:,\;\;\;\;
  e_2 \ot e_2 \:,
\]
while the~eigenvalue~$-q^2$ corresponds to an~antisymmetric vector
\[
  e_1 \ot e_2 - q \, e_2 \ot e_1 \:.
\]
Let us define intertwiners
\begin{equation}
  \hlabel{c:si-k}
  \si_k =
  \underbrace{\id \ot \ldots \ot \id}_{\mbox{\scriptsize $k-1$ \ times}}
                                                        \ot \; \si \ot
  \underbrace{\id \ot \ldots \ot \id}_{\mbox{\scriptsize $M-k-1$ \ times}}
  \;, \;\;\;\;   k=1,2, \ldots, M-1
\end{equation}
acting~in $K^{\ot M}$ ($M \in \N$).
This definition and the~properties (\ref{c:si2})
and~(\ref{c:si1}) of~$\si$ imply the~relations of Hecke algebra:
\renewcommand{\arraystretch}{1.3}%
\begin{equation}
\hlabel{c:warkocz}
\begin{array}{cc}
  \left.
  \begin{array}{l}
    \si_k \si_l = \si_l \si_k \:, \\
    \si_k \si_{k+1} \si_k = \si_{k+1} \si_k \si_{k+1} \:, \\
    {\si_k}^2 = (1-q^2) \si_k + q^2
  \end{array}
  \right\}
  &
  \begin{array}{l}
    |k-l| \geq 2 \:, \\
    k,l=1,\ldots, M-1 \:. \\
  \end{array}
\end{array}
\end{equation}
\renewcommand{\arraystretch}{1.3}%

Let
\begin{eqnarray}
  \hlabel{c:Ekdef}
  &  E_k  =
  \underbrace{\id \ot \ldots \ot \id}_{
         \mbox{\scriptsize $k-1$ \ times}} \ot E \ot
  \underbrace{\id \ot \ldots \ot \id}_{
         \mbox{\scriptsize $M-k-1$ \ times}}
                                   \;, \;\;\;  k=1,2, \ldots, M-1\;.  \\
  \hlabel{c:Ek'def}
  &  E'_k =
  \underbrace{\id \ot \ldots \ot \id}_{
         \mbox{\scriptsize $k-1$ \ times}} \ot E' \ot
  \underbrace{\id \ot \ldots \ot \id}_{
         \mbox{\scriptsize $M-k-1$ \ times}}
                                   \;, \;\;\;  k=1,2, \ldots, M-1\;.
\end{eqnarray}
Let us define a~subspace of symmetric vectors
\begin{equation}
  \hlabel{c:wektsym1}
  K^{\frac{M}{2}} = K^{\ot_{\rm sym}\,M} =
     \{x \in K^{\ot M} \,:\,
           \si_k x = x \:, \;\;\; k=1,\ldots,M-1 \} \:.
\end{equation}
The~intertwiner $E_k$ is an~injection and therefore
\begin{equation}
  \hlabel{c:wektsym2}
  K^{\frac{M}{2}} =
     \{x \in K^{\ot M} : E'_k x = 0 \:, \;\;\; k=1, \ldots, M-1 \} \:.
\end{equation}
$K^{\frac{M}{2}}$ is equal to the~intersection
of kernels of intertwiners~$E'_k$, $k=1,\ldots, M-1$,
hence it is an~invariant subspace of $K^{\ot M}$.
Thus one can define the~subrepresentation
\begin{equation}
  \hlabel{c:uNdef}
  u^{\frac{M}{2}} = \restr{ u^{\tps M} }{ K^{\frac{M}{2}} }  \:.
\end{equation}
In~particular, $u^0=I$, $u^{\jnd}=u$.

Each permutation $\pi \in \Pi(M)$ can be written as follows
\begin{equation}
  \hlabel{c:Tperm}
  \pi = t_{i_1} \ldots t_{i_m} \:, \;\;\;
    \begin{array}{l}
     1 \leq i_1, \ldots, i_m \leq M-1  \:, \\
     i_1,\ldots,i_m \in \N \:,
    \end{array}
\end{equation}
where $t_j = (j,j+1)$ is a~transposition and $m=m(\pi)$ is minimal.
It is known that $m(\pi)$ is equal to the~number of $\pi$-inversions.
The~intertwiner
\begin{equation}
  \hlabel{c:sipidef}
  \si_{\pi}=\si_{i_1} \ldots \si_{i_m}
\end{equation}
does not depend on a~choice of a~minimal
decomposition~(\ref{c:Tperm}), which can be obtained from (\ref{c:warkocz})
(cf~\cite{b:Differential}, page 154).

Let us define an~operator of $q$-symmetrization
$S_M : K^{\ot M} \rightarrow K^{\ot M}$ given~by (cf. \cite{b:Jimbo})
\begin{equation}
  \hlabel{c:SN}
  S_M  = \sum_{ \pi \in \Pi(M) } q^{-2 \iInv{\pi}} \si_{\pi} \:.
\end{equation}

\begin{fakt}
\hlabel{c:si-id}
$ ( \si_k - \id ) S_M = 0 $ \ for \ $k=1,\ldots, M-1$.
\end{fakt}

\dow
Let us call a~permutation $\pi \in \Pi(M)$ ''good'', if
$\pi^{-1}(k) < \pi^{-1}(k+1)$ for fixed~$k$. A~''bad'' permutation
is meant to be not a~''good'' one.
If $ \pi = t_{i_1} \ldots t_{i_l} $ is a~minimal decomposition of
a~''good'' permutation into transpositions, then a~minimal decomposition
of the~''bad'' permutation $ t_k \pi $ is
$ t_k t_{i_1} \ldots t_{i_l} $
and $\iINV{t_k \pi}=\iINV{\pi}+1$, $\si_{t_k \pi}=\si_k \si_{\pi}$.
In such a~way ''good'' and ''bad''
permutations correspond bijectively.
According to~(\ref{c:SN}), one has
\renewcommand{\arraystretch}{0.4}%
\begin{eqnarray*}
  S_M & = & \hspace{1em}
  \underbrace{ \sum_{
    \begin{array}{c}
      \scriptstyle \pi \in \Pi(M) \\
      \scriptstyle \pi \mbox{ \scriptsize is ''good''}
    \end{array}
    }
    q^{-2 \iInv{\pi}} \si_{\pi}
  }_{
    \begin{array}{c}
       \rule[-0.9ex]{0in}{0in} \scriptscriptstyle \parallel \\
       \scriptstyle S_M^{(0)}
    \end{array}
  } \hspace{1.5em} +
  \sum_{
    \begin{array}{c}
      \scriptstyle \pi' \in \Pi(M) \\
      \scriptstyle \pi' \mbox{ \scriptsize is ''bad''}
    \end{array}
  }
  q^{-2 \iInv{\pi'} } \si_{\pi'}    \\ \\
& = & ( \id + q^{-2} \si_k ) \: S_M^{(0)} \;,
\end{eqnarray*}
\renewcommand{\arraystretch}{1}%
where $S_M^{(0)}$ is defined in the~first line of the~formula.
Using~(\ref{c:warkocz}) one can check the~equation
$ \si_k S_M = S_M $.
\cnd

\begin{wniosek}
  \hlabel{c:imageSN}
  $\image S_M \subset K^{\frac{M}{2}}$.
\end{wniosek}

Let us define $\fact_x$ for $x \in \C$ as follows
\begin{equation}
  \hlabel{c:factdef}
  \fact_x(M) = \sum_{ \pi \in \Pi(M) } x^{ 2 \iInv{\pi} } \:.
\end{equation}

Using the~mathematical induction one can prove
\begin{fakt}
\hlabel{c:factfakt}
$
  \fact_x(M) = \left\{
    \begin{array}{cl}
       \renewcommand{\arraystretch}{0.8}
         \begin{array}{c}
           \scriptstyle M \\
           \prod \\
           \scriptstyle k=1
         \end{array}
       \renewcommand{\arraystretch}{1}
       \frac{1-x^{2 k}}{1-x^2} \;, & x \neq \pm 1 \\
       \rule{0in}{3ex}  M! \;, & x = \pm 1 \;.
    \end{array}
  \right.
$
\end{fakt}

\begin{wniosek}
  \hlabel{c:FactOnRoots}
  $ \fact_{\frac{1}{q}}(M) = 0 $ if and only if $ M \geq N_0 $.
\end{wniosek}

Using~(\ref{c:SN}), Proposition~\ref{c:si-id}.\ and~(\ref{c:factdef}),
one can obtain
\begin{fakt}
  \hlabel{c:SN'2}
  $S_M^2 = \fact_{\frac{1}{q}}(M) S_M$ .
\end{fakt}

\begin{fakt}
\hlabel{c:dimKN-2}
  $\dim K^{\frac{M}{2}} = M+1$.
\end{fakt}

\dow
For a~given element $x \in K^{\ot M}$ one can write a~decomposition
\begin{equation}
  \hlabel{c:xrozklad}
  x = \sum_{i_1, \ldots, i_M = 1,2} x_{i_1, \ldots, i_M}
  e_{i_1} \ot \ldots \ot e_{i_M} \;,
\end{equation}
where $ x_{i_1, \ldots, i_M} \in \C $.
According to~(\ref{c:wektsym2}), the~statement $x \in K^{\frac{M}{2}}$ is
equivalent to: $\forall k=1,2,\ldots,M-1$ \ $E'_k x = 0$,
which can be replaced by: for all $k=1,2,\ldots,M-1$ and for all
$x_{i_1, \ldots, i_M}$ such that $i_k=1$ and $i_{k+1}=2$ the~following holds
\begin{equation}
  \hlabel{c:xwar}
  x {\raise-2pt\hbox{$\scriptstyle \ldots
   \raise0.5pt\hbox{$\scriptstyle\stackrel{\stackrel{k}{\vee}}{1}$}2
   \ldots$ } }
  = q\, x {\raise-2pt\hbox{$\scriptstyle \ldots
   \raise0.5pt\hbox{$\scriptstyle\stackrel{\stackrel{k}{\vee}}{2}$}1
   \ldots$ } } \:,
\end{equation}
where $ \scriptstyle \stackrel{k}{\vee} $ denotes the~$k$--th position
of an~index.
It means that all the~coefficients $ x_{i_1, \ldots, i_M} $
can be uniquely computed from the~coefficients
$ x_{1 \ldots 1} $, $x_{1 \ldots 12}$, $\ldots$, $x_{2 \ldots 2} $.
One can conclude now that the~thesis holds.
\cnd

\begin{Uwaga}
\hlabel{c:SIKprzestawienie}
  The~linear span $W_l$ of
  $e_{j_1} \ot \ldots \ot e_{j_M}$, \ $j_1,\ldots,j_M=1,2$
  with a~given number~$l$ of $m$ such that $j_m=1$, is invariant
  w.r.t. 
  the~intertwining operators $\si_k$, $k=1,2,\ldots,M-1$ as well as
  w.r.t. 
  the~operator~$S_M$.
\end{Uwaga}

The~above remark can be directly obtained from (\ref{c:si-k}),
(\ref{c:si()}) and the~definition~(\ref{c:SN}) of~$S_M$.

\begin{fakt}
  \hlabel{c:dimImSNg}
  The~following inequalities hold
  \begin{enumerate}
    \item $
       \dim \image S_M \geq M + 1$ \ \ for $M \in \N$ such that $M < N_0$,
    \item $
       \dim \image S_{N_0} \geq N_0 - 1$ (for $N_0 < \infty$).
  \end{enumerate}
\end{fakt}

\dow
Let us fix~$M$. Using Remark~\ref{c:SIKprzestawienie}.\ one can see
that in a~decomposition of
\begin{equation}
  \hlabel{c:SNe}
  S_M(
  \underbrace{ e_1 \ot \ldots \ot e_1 }_{\mbox{\scriptsize $k$ times}} \ot
  \underbrace{ e_2 \ot \ldots \ot e_2 }_{\mbox{\scriptsize $M-k$ times}}
  )\:,
  \;\;\;\;  k=0, \ldots, M
\end{equation}
there are only the~elements of the~basis $ e_{i_1} \ot \ldots \ot e_{i_M} $
that have the~number of $e_1$ equal to~$k$.

It can be easily computed that an~element
$
  \underbrace{ e_1 \ot \ldots \ot e_1 }_{\mbox{\scriptsize $k$ times}} \ot
  \underbrace{ e_2 \ot \ldots \ot e_2 }_{\mbox{\scriptsize $M-k$ times}}
$
has the~coefficient equal~to
\[
   \fact_{\frac{1}{q}} (k) \; \fact_{\frac{1}{q}} (M-k) \:.
\]
For $M<N_0$ all these coefficients are nonzero and
the~elements~(\ref{c:SNe}) are linearly independent.
In the~case $M=N_0$ one gets coefficients equal~$0$ only for $k=0$
and $k=N_0$.
\cnd

Using Corollary~\ref{c:imageSN}.,
Proposition~\ref{c:dimImSNg}.\ and Proposition~\ref{c:dimKN-2}., we get
\begin{wniosek}
  \hlabel{c:dimImSN}
  If $M$ is a~natural number such that $M < N_0$ then
  \[ \begin{array}{c}
    \dim \image S_M = M+1 \:,  \\
    \image S_M = K^{\frac{M}{2}} \:.
  \end{array} \]
\end{wniosek}

\begin{lemat}
\hlabel{c:philemat}
Let $\varphi$ be the~intertwiner defined by
  \begin{eqnarray*}
     &   \varphi : K^{s-\jnd} \longrightarrow K^s \ot K \:, \\
     &   \varphi( x ) = (S_{2 s} \ot \id )(x \ot E(1) ) \:,
         \;\;\; x \in K^{s-\jnd} \:,
  \end{eqnarray*}
for given $s=0,\jnd,1,\ldots$ such that $2 s < N_0 - 1$.
Moreover, let the~representation $u^{s-\jnd}$ be irreducible.
Then
  \begin{enumerate}
    \item $\ker \varphi = \{0\}$,
    \item $\image \varphi$ corresponds to the~representation $u^{s-\jnd}$,
    \item $\image \varphi \cap K^{s+\jnd} = \{0\}$.
  \end{enumerate}
\end{lemat}

\dow
We compute
\begin{eqnarray*}
  \lefteqn{
    \varphi( e_1 \ot \ldots \ot e_1 )
  } \\
  & = &  S_{2s}( e_1 \ot \ldots \ot e_1 ) \ot e_2 -
          q \, S_{2s}( e_1 \ot \ldots \ot e_1 \ot e_2 ) \ot e_1 \\
  & = &  \fact_{\frac{1}{q}}( 2s ) \, e_1 \ot \ldots \ot
                                     \underline{e_1 \ot e_2} -  \\
  & & \hspace{4em} q \, \left[ \, \fact_{\frac{1}{q}}( 2s -1 ) \,
                    e_1 \ot \ldots \ot e_1 \ot \underline{e_2 \ot e_1} +
      \ldots \, \right]  \neq 0 \:.
\end{eqnarray*}
Thus irreducibility of~$u^{s-\jnd}$ implies~1.
The~result of~2.\ is now obvious.

$\image \varphi $ corresponds to the~irreducible representation
$u^{s-\jnd}$. Thus if 3.\ would not hold then
$\image \varphi \Subrep K^{s+\jnd}$,
because $K^{s+\jnd}$ is $u^s \tp u$ invariant (see~(\ref{c:wektsym2})).
This implies $\varphi( e_1 \ot \ldots \ot e_1 ) \in K^{s+\jnd}$.
Applying~(\ref{c:xrozklad}) and (\ref{c:xwar}) to
the~underlined elements of $x=\varphi(e_1 \ot \ldots \ot e_1)$
one gets
\[
  \fact_{\frac{1}{q}}(2s) = q \; (-q) \fact_{\frac{1}{q}}(2s-1) \:,
\]
which is impossible (see Proposition~\ref{c:factfakt}).
This contradiction shows~3.
\cnd

\begin{fakt}
\hlabel{c:urozklad}
Let the~representations $u^0$, $u^{\jnd} , \ldots , u^s$ be
irreducible for fixed $s \in \frac{\N}{2}$ such that
$\jnd \leq s < \frac{N_0}{2}-\jnd$. Then
\begin{enumerate}
  \item
     \[  u^s \tp u \approx u^{s-\jnd} \oplus u^{s+\jnd} \:,  \]
  \item the~representation $u^{s+\jnd}$ is irreducible.
\end{enumerate}
\end{fakt}

\dow
1. Using (\ref{c:wektsym2}) one has
\[
  K^{s+\jnd} \Subrep K^s \ot K \:.
\]
Applying the~intertwiner~$\varphi$ of Lemma~\ref{c:philemat}.\ one gets
\begin{equation}
  \hlabel{c:philemmatemp}
  K^{s +\jnd} \oplus \image \varphi \; \Subrep \; K^s \ot K \:.
\end{equation}
The~dimensions of both sides of the~inclusion are the~same. This
proves~1.

2. Analogously as in~1.\ one has
$u^i \tp u \approx u^{i-\jnd} \oplus u^{i+\jnd}$,
$i = \jnd, 1, \jnd, \ldots, s$.
Using the~mathematical induction,
$u^{\tps l}$ can be decomposed into a~direct sum of some copies of
$ u^{\frac{l}{2}}$, $u^{\frac{l}{2}-1}$, $\ldots$, $u^{\jnd}$ or~$u^0$,
$l=0,1,\ldots,2s+1$.

Therefore
\renewcommand{\arraystretch}{0.3}%
\begin{eqnarray}
  \lefteqn{
    \dim A_{2s+1}
  } \nonumber \\
  & = & \dim \: \spn \left\{ ( u^{\tps l} )_{k m} \, : \,
      l=0,1,\ldots, 2s+1 \:, \;\; k,m=1,\ldots, 2^l \right\}
      \nonumber \\
\hlabel{c:dimu}
   & \leq  & \dim \: \spn \left\{ u^{\frac{l}{2}}_{km} \, : \,
       l=0,1,\ldots, 2s+1 \:, \;\; k,m=1,\ldots, (l+1) \right\}
\end{eqnarray}
\renewcommand{\arraystretch}{1}%
Comparing this with~(\ref{c:Akdim}) one obtains that the~matrix elements
of the~representation $u^{s+\jnd}$ must be linearly independent
and the~representation $u^{s+\jnd}$ must be irreducible.
\cnd

Using the~last proposition and the~mathematical induction one can
prove
\begin{twierdzenie}
  \hlabel{c:twUrozklad}
  The~representations $u^s$,
  $s \in \frac{\N}{2}$, $s < \frac{N_0}{2}$
  are irreducible and the~following decomposition holds
  \[
    u^s \tp u \approx u^{s-\jnd} \oplus u^{s+\jnd} \:,
    \;\;\; s < \frac{N_0}{2} - \jnd.
  \]
\end{twierdzenie}

\begin{Uwaga}
  \hlabel{c:SomeCopiesOfU}
  In particular the~representation~$u^{\tps M}$
  is a~direct sum of some copies of representations
  $u^{\frac{M}{2}}$, $u^{\frac{M}{2}-1}$, $\ldots$,
  $u^{\jnd}$ or $u^0$ for $M \in \N$ such that $M < N_0$.
\end{Uwaga}

Using the~above theorem and Proposition~A.2.\
in~\cite{b:CQGandTRR} one gets
\begin{wniosek}
  \hlabel{c:allreducible}
  Each representation of $SL_q(2)$ is a direct sum of some copies of
 $u^s$, $s \in \N/2$,  for
  $q \in \C \setminus \{0, \mbox{roots of unity}\}$.
\end{wniosek}


\hsection{The~case of roots of unity}
\hlabel{m:maincase}

The~complex number $q$ is assumed to be a~root of unity all over the~section.
In this case $N_0$ is finite (see section~\ref{i:introduction})
and $N_0 \geq 2$.

\hsubsection{The~basic decomposition}
\hlabel{m:BasicDecomposition}

   From the~proof of Theorem~\ref{c:twUrozklad}.\ one can see that
a~decomposition of $u^{\frac{N_0}{2}-\jnd} \tp u$ may be completely
different from the~one in that Theorem. \\
The~aim of the~present subsection is to find it.

\vspace{3ex}

Let $L$ be the~subspace of $K^{\ot N_0}$ given by the~formula
\begin{equation}
  \hlabel{m:Ldef}
  L = \spn \left\{
                e_k \ot \ldots \ot e_k \in K^{\ot N_0} : k = 1,2
           \right\}   \:.
\end{equation}
One can see that $\dim L = 2$ and $L \Subrep K^{\frac{N_0}{2}}$.

\begin{lemat}
  \hlabel{m:philemat}
  Let $\varphi$ be the~following intertwiner
    \begin{eqnarray*}
       &   \varphi : K^{\frac{N_0}{2}-1} \longrightarrow
                     K^{\frac{N_0}{2}-\jnd} \ot K \:, \\
       &   \varphi( x ) = (S_{N_0-1} \ot \id )(x \ot E(1) )
    \end{eqnarray*}
  for $x \in K^{\frac{N_0}{2}-1}$.
  Then
    \begin{enumerate}
      \item  $\ker \varphi = \{0\}$,
      \item  $\image \varphi$ corresponds to the~irreducible representation
             $u^{\frac{N_0}{2}-1}$,
      \item  $ K^{\frac{N_0}{2}} = \image \varphi \oplus L $.
    \end{enumerate}
\end{lemat}

\dow
  1.\ and 2.\ \ See Lemma~\ref{c:philemat}.

  3. One can easily check
  (in different manners for $k=1,2,\ldots, N_0-2$ and for $k=N_0-1$)
  that
  \[
    E'_k \varphi( e_1 \ot \ldots \ot e_1 ) = 0
    \mbox{ for } k=1,2,\ldots, N_0-1 \:.
  \]
  It means that
  \begin{equation}
    \hlabel{m:varphi}
    \varphi( e_1 \ot \ldots \ot e_1 ) \in K^{\frac{N_0}{2}} \:.
  \end{equation}
  Moreover, the~above element is different from zero.

  According to~2.\
  $\image \varphi$ corresponds to an~irreducible representation
  and therefore
  \begin{equation}
    \hlabel{m:phisubK}
    \image \varphi \; \Subrep \; K^{\frac{N_0}{2}} \:.
  \end{equation}

  Decomposing an~element of~$\image \varphi$ into elements of
  the~basis $e_{i_1} \ot \ldots \ot e_{i_{N_0}}$, \ \
  $i_1, \ldots, i_{N_0} = 1,2$,
  one can see that the~elements $e_k \ot \ldots \ot e_k$, $k=1,2$
  have the~coefficients equal~$0$ (we use Remark~\ref{c:SIKprzestawienie}).
  Thus
  \[
    \image \varphi \cap L = \{ 0 \} \:.
  \]
  Calculating the~dimensions one can prove
  \[
    K^{\frac{N_0}{2}} = \image \varphi \oplus L  \:.
  \]
\cnd

\begin{Uwaga}
  Let $\hat{S}_{M}$, $M \in \N$ be an~intertwiner defined as follows
  \begin{eqnarray*}
    & \hat{S}_{M} = \restr{ S_{M} }{ K^{\frac{M}{2}-\jnd} \ot K } \\
    & \hat{S}_{M} :
      K^{\frac{M}{2} - \jnd} \ot K \longrightarrow K^{\frac{M}{2}}
  \end{eqnarray*}
  One can prove
  \[
    \image \varphi = \image \hat{S}_{N_0} =
    \image S_{N_0} \subset K^{\frac{N_0}{2}} =
    \ker \hat{S}_{N_0} \subset \ker S_{N_0} \:,
  \]
  which gives (cf Proposition~\ref{c:SN'2}) the~equalities
  $\hat{S}_{N_0}^2 = 0$, $S_{N_0}^2 = 0$. The~situation for $M<N_0$ was
  completely distinct: $S_M$ was proportional to a~projection and hence
  \renewcommand{\arraystretch}{1.3}%
  \[
    \begin{array}{c}
      \image S_M \oplus \ker S_M = K^{\otimes M} \:, \\
      \image \hat{S}_M \oplus \ker \hat{S}_M =
           K^{\frac{M}{2}-\jnd} \otimes K \:.
    \end{array}
  \]
  \renewcommand{\arraystretch}{1}%
\end{Uwaga}

\begin{fakt}
  \hlabel{m:Lquotientlemma}
  Let
  \begin{equation}
    \hlabel{m:vdef}
    v= \left(
          \matrix{ \al^{N_0} & \be^{N_0} \cr \ga^{N_0} & \de^{N_0} \cr }
       \right) \:.
  \end{equation}
  Then $v$ is a~quotient irreducible representation of a~subrepresentation of
  \\
  $u^{\frac{N_0}{2}-\jnd} \tp u$, corresponding to the~quotient space
  $\quot{K^{\frac{N_0}{2}}}{\image \varphi} \approx L$.
\end{fakt}

\dowspace
\dow
  In virtue of~(\ref{c:uNdef}) $K^{\frac{N_0}{2}}$ is
  $u^{\frac{N_0}{2}-\jnd} \tp u$-invariant subspace.

  Using Lemma~\ref{m:philemat}.\ one can see that
  $\quot{K^{\frac{N_0}{2}}}{\image \varphi} \approx L$ corresponds to
  a~representation.
  Its matrix elements are the~matrix elements of the~representation
  $u^{\tps N_0}$ that
  appear at the~intersections of columns and rows corresponding to
  $e_1 \ot \ldots \ot e_1$ and $e_2 \ot \ldots \ot e_2$
  and are given by~(\ref{m:vdef}).
  The~elements of~$v$ are elements of the~basis~(\ref{c:elembazy})
  of the~algebra~$A$ and therefore $v$ is irreducible.
\cnd

\begin{lemat}
  \hlabel{m:E'lemat}
  Let us define an~intertwiner
  \begin{eqnarray*}
    &  \hat{E}'_{m-1} =
        \restr{ E'_{m-1} }{ K^{\frac{m}{2}-\jnd} \ot K } \:, \\
    &  \hat{E}'_{m-1} : K^{\frac{m}{2}-\jnd} \ot K \longrightarrow
                                               K^{\ot (m-2)} \:,
  \end{eqnarray*}
  $m = 2,3,\ldots$. Then
  \begin{enumerate}
    \item $ \ker \hat{E}'_{m-1} = K^{\frac{m}{2}} $,
    \item $ \image \hat{E}'_{m-1} = K^{\frac{m}{2}-1}$,
    \item $ \image \hat{E}'_{m-1}$ corresponds to the~representation
                                            $u^{\frac{m}{2}-1}$.
  \end{enumerate}
\end{lemat}

\dow
  1.\ follows from~(\ref{c:wektsym2}) and the~definition of
  $\hat{E}'_{m-1}$.

  2. One can see (cf~\ref{c:xwar}) that
  \begin{equation}
    \hlabel{m:temphat}
    \image \hat{E}'_{m-1} \subset K^{\frac{m}{2}-1} \:.
  \end{equation}
  One has
  \[
    \dim ( K^{\frac{m}{2}-\jnd} \ot K ) =
         \dim \ker \hat{E}'_{m-1} + \dim \image \hat{E}'_{m-1} \:.
  \]
  Thus    $ \dim \image \hat{E}'_{m-1} = m - 1 $
  (see Proposition~\ref{c:dimKN-2}.).
  That and~(\ref{m:temphat}) prove~2.

  3.\ follows from~(\ref{c:uNdef}).
\cnd

\begin{wniosek}
  \hlabel{m:uTdecomposition}
  The~following decomposition holds (cf Appendix~\ref{q:qkor}.)
  \[
    u^s \tpt u \approx \widetilde{u^{s-\jnd}} \oplus \widetilde{u^{s+\jnd}}
    \;\;\; \mbox{for $s = \jnd,1,\ldots$. }
  \]
\end{wniosek}

\dow
  We notice
  $\quot{ K^{\frac{m}{2}-\jnd} \ot K }{
          \ker \hat{E}'_{m-1} } \approx \image \hat{E}'_{m-1}$ and set
  $s=\frac{m}{2}-\jnd$.
\cnd

\begin{wniosek}
  \hlabel{m:uQuotWn}
  The~quotient space
  $\quot{ ( K^{\frac{N_0}{2}-\jnd} \ot K ) }{ K^{\frac{N_0}{2}} }$
  corresponds to the
  representation $u^{\frac{N_0}{2}-1}$, which is irreducible.
\end{wniosek}

\begin{twierdzenie}
  \hlabel{m:twUVrozklad}
  \begin{enumerate}
    \item
    \renewcommand{\arraystretch}{1.4}%
    $
     u^{\frac{N_0}{2}-\jnd} \tp u \approx
     \left(
     \begin{array}{c|c|c}
        u^{\frac{N_0}{2}-1} & \ast & \ast                \\ \hline
        0                   & v    & \ast                \\ \hline
        0                   & 0    & u^{\frac{N_0}{2}-1}
     \end{array}
     \right)
    $.
    \renewcommand{\arraystretch}{1}%
    \item $ \widetilde{u^{\frac{N_0}{2}}}
            \approx u^{\frac{N_0}{2}-1} \oplus v $.
    \item
    All elements denoted by three stars are linearly independent
    from each other as well as from the~matrix elements of
    the~representations $u^{\frac{N_0}{2}-1}$ and~$v$.
  \end{enumerate}
\end{twierdzenie}

\dow
  1.\ and 2. We use Lemma~\ref{m:philemat}.,
  Proposition~\ref{m:Lquotientlemma}.\ and Corollary~\ref{m:uQuotWn}.

  3. Using Theorem~\ref{c:twUrozklad}.\
  one can see that the~representation
  $u^{\tpstil N_0}$
  decomposes into some copies of $u^{\frac{N_0}{2}-\jnd} \tp u$,
  $u^{\frac{N_0}{2}-1}$, $u^{\frac{N_0}{2}-2}$, $\ldots$, $u^{\jnd}$ or $u^0$.

  The~following is obvious
  \[
    A_{N_0} = \spn \left\{ (u^{\tps k})_{ij} \; : \; k=0,1,\ldots,N_0, \;\;\;
              i,j=1,2,\ldots,2^k \right\} \:,
  \]
  where $A_k$ is defined by~(\ref{c:Akdef}).

  According to Remark~\ref{c:SomeCopiesOfU}.\ and
  1.~of~Theorem~\ref{m:twUVrozklad}.\ one has
  \[
    A_{N_0} \; = \;
    \spn \left\{
      \left(
        \parbox{2.5cm}{\small elements denoted by three stars}
      \right) \:,
      v_{i'j'}, \; u^{\frac{k}{2}}_{ij} \; : \;
      \begin{array}{c}
        i',j'=1,2,      \\
        k=0,1,2,\ldots,N_0-1, \\
        i,j =1,2,\ldots,k+1
      \end{array}
    \right\} \:.
  \]
  Comparing the~dimensions of both sides one gets~3.~(see~(\ref{c:Akdim})).
  \nopagebreak \\
\cnd


\hsubsection{Irreducible representations}
\hlabel{m:irreducible}

\begin{fakt}
  \hlabel{m:algebra'fakt}
  Let $A'$ be the~subalgebra of $A$ generated by the~elements
  \[
    \al' = \al^{N_0} \:,\;\; \be' = \be^{N_0} \:,\;\;
    \ga' = \ga^{N_0} \:,\;\; \de' = \de^{N_0} \:.
  \]
  Then $A'$ is isomorphic to the~algebra $A$ for the~changed parameter
  $q' = q^{{N_0}^2}$.
\end{fakt}

\uwaga
The~new parameter $q'$ may be equal~$\pm 1$.

\dowspace
\dow
  Using Proposition~\ref{c:baza}.\ one can see that the~elements
  $\al'_k \ga'^m \be'^n$, $ k \in \Z $, \ $ m,n \in \N $, \
  are linearly independent in the~algebra~$A'$,
  where $\al'_k$ are defined in analogous way as $\al_k$ (see~(\ref{c:alk})).

  It suffices to prove that the~elements $\al'$, $\be'$, $\ga'$, $\de'$
  fulfill the~relations~(\ref{c:relations}) for the~new parameter~$q'$.
  The~first five relations are immediate to prove,
  the~last two will be considered now.

  One has the~following equation of polynomials
  \begin{equation}
    \hlabel{m:KacEquat}
    \prod_{j=0}^{N_0-1} ( 1 + q^{2j} x ) = 1 + q^{N_0 (N_0-1)} x^{N_0} \:.
  \end{equation}

  Using the~mathematical induction one can obtain
  \[
    \al^{N_0} \de^{N_0} = \prod_{j=0}^{N_0-1} (1 + q^{2j} (q \be \ga)) \:,
  \]
  which using~(\ref{m:KacEquat}) can be written~as
  \[
    \al^{N_0} \de^{N_0} = I + q^{N_0^2} \be^{N_0} \ga^{N_0} \:.
  \]
  This corresponds to the~last but one relation of~(\ref{c:relations}).
  The~last relation can be proved in a~similar way.
\cnd

\begin{Uwaga}
\hlabel{m:CentrumA}
  One can easily check that $A'$ is contained in the~center of~$A$
  for $q$ being a~root of unity of an~odd degree.
\end{Uwaga}

\vspace{2ex}
Using the~notations introduced in Proposition~\ref{m:algebra'fakt}.,
representations $v^s$, $s=0,\jnd,1, \ldots$, can be defined in an~analogous
way as the~representations $u^s$, $s=0,\jnd,1, \ldots$,
(but for the~parameter $q' = q^{N_0^2}$, see (\ref{c:uNdef})).

According to Theorem~\ref{c:twUrozklad}.\ one has
($N_0=+\infty$ for $q=\pm 1$)
\begin{wniosek}
  \hlabel{m:twVrozklad}
  The~representations $v^s$, $s=0,\jnd,1, \ldots$ are irreducible.
  Moreover the~following decomposition holds
  \[
    v^s \tp v \approx v^{s-\jnd} \oplus v^{s+\jnd} \:.
  \]
\end{wniosek}

\begin{fakt}
  \hlabel{m:NPrep}
  The~representations
  \[
    v^t \tp u^s \:, \;\;\;  t=0,\jnd,1,\ldots \:, \;\;\;
                        s=0,\jnd,1,\ldots,\frac{N_0}{2}-\jnd
  \]
  are irreducible and nonequivalent.
\end{fakt}

\dow
  Is suffices to prove linear independence of matrix elements of
  representations
  $v^t \tp u^s$, \
  $t=0,\jnd,1,\ldots$, $s=0,\jnd,1,\ldots,\frac{N_0}{2}-\jnd$.
  But matrix elements of $v^t \tp u^s$ belong to
  $A_{N_0 2 t + 2 s}$ and the~numbers $N_0 2 t + 2 s$ are different for
  different pairs $(t,s)$. Therefore we need to prove (for given $t$, $s$)
  the~linear independence modulo $A_{N_0 2 t + 2 s - 1}$
  of the~matrix elements
  \[
    \left( v^t \tp u^s \right)_{ik,jl} =
    v^t_{ij} u^s_{kl} \:, \;\;\;
    i,j = 1, 2, \ldots, 2 t \;\; k,l = 1, 2, \ldots, 2 s
  \]
  which is equivalent to the~linear independence
  (in our sense) of the~elements
  \begin{equation}
    \hlabel{m:baza'baza}
    \al_{N_0 a} \be^{N_0 b} \ga^{N_0 c} \al_k \be^m \ga^n \:,
  \end{equation}
  where
  $a,k \in \Z$, \ $b,c,m,n \in \N$ are such that
  $|a| + b + c = 2 t$ and $|k| + m + n = 2 s$
  (matrix elements of $u^s$ modulo $A_{2s-1}$
  are basis of $\quot{A_{2s}}{A_{2s-1}}$,
  cf~(\ref{c:dimu}), similarly for $v^t$).

  Doing some computations one can see that the~linear independence of
  the~elements~(\ref{m:baza'baza}) is equivalent to the~linear independence
  of the~elements
  \[
  \renewcommand{\arraystretch}{1.3}
    \begin{array}{lll}
      \mbox{(1)} &
      (\al^{N_0})^a (\be^{N_0})^b (\ga^{N_0})^c \: \al^k \be^m \ga^n \:, \\
      \mbox{(2)} &
      (\de^{N_0})^a (\be^{N_0})^b (\ga^{N_0})^c \: \de^k \be^m \ga^n \:, &
      \mbox{$a \geq 1$ or $k \geq 1$,}   \\
      \mbox{(3)} &
      (\al^{N_0})^{a-1} (\be^{N_0})^b (\ga^{N_0})^c
                          \: \al^{N_0-k} \be^{m+k} \ga^{n+k} \:, &
      a,k \geq 1 \:, \\
      \mbox{(4)} &
      (\de^{N_0})^{a-1} (\be^{N_0})^b (\ga^{N_0})^c
                          \: \de^{N_0-k} \be^{m+k} \ga^{n+k} \:, &
      a,k \geq 1 \:,
    \end{array}
  \renewcommand{\arraystretch}{1}
  \]
  where $a,b,c,k,m,n \in \N$ are such that
  $a + b + c = 2 t$ and $k + m + n = 2 s$.
  The~above elements are proportional to some elements of
  the~basis (\ref{c:elembazy}).

  It can be easily seen that the~elements (1) and~(3) are of a~different
  form than the~elements (2) and~(4).

  Moreover the~elements may be characterized by the~number
  being the~sum of the~powers of the~elements:
  $\be$, $\ga$ and $\al$ (or $\de$), where we do not take
  into consideration $\al^{N_0}$, $\be^{N_0}$, $\ga^{N_0}$, $\de^{N_0}$.
  The~number $2 s$ corresponds to the~elements (1) and~(2),
  the~number $N_0 + 2 s$ corresponds to the~elements (3) and~(4).
  Remember that $2 s < N_0$.

  The~elements (1), (2), (3) and~(4) have the~degree $N_0 2 t + 2 s$,
  hence they are independent modulo $A_{N_0 2 t + 2 s - 1}$
  and the~proof is finished.
\cnd


\hsubsection{All irreducible representations}
\hlabel{i:allirreducible}

Let
\[
  A_{\rm cent} = \left\{ a \in A \; : \; \flip(\Kom(a)) = \Kom(a) \right\} \:,
\]
where $\flip : A \ot A \rightarrow A \ot A$ is a linear mapping such that
$\flip(a \ot b) = b \ot a$ for all $a,b \in A$, \ $\Kom$ is
the~comultiplication.
\begin{lemat}
  \hlabel{i:Acent}
\[
 A_{\rm cent} =
    \spn \left\{ \tr ( v^t \tp u^s ) \; : \; t=0,\jnd,1,\ldots \:,\;\;
                                   s=0,\jnd,1,\ldots,\frac{N_0}{2}-\jnd
                \right\} \:.
\]
\end{lemat}
Using the~above lemma, the~fact that traces of all irreducible
representations form a~linearly independent subset in
$A_{\rm cent}$ and Proposition~\ref{m:NPrep}.,
one obtains
\begin{twierdzenie}
  The~representations
  \[
    v^t \tp u^s, \hspace{1cm}
       t=0,\jnd,1,\ldots, \hspace{0.5cm}
       s=0,\jnd,1,\ldots,\frac{N_0}{2}-\jnd
  \]
  are all nonequivalent irreducible representations of
  the~quantum group $(A,u)$ for $q$ being a~root of unity.
\end{twierdzenie}
Note that for $q=\pm 1$ the~above theorem is also true. In this case one
could put $N_0=1$ and~$u=v$.
\dowspace
\refdow{of Lemma~\ref{i:Acent}.}
  Due to Corollary~\ref{m:twVrozklad}., Theorem~\ref{m:twUVrozklad}.\
  and Theorem~\ref{c:twUrozklad}.,
  \renewcommand{\arraystretch}{1.4}%
  \begin{equation}
  \hlabel{i:tracesUV}
    \begin{array}{r @{\hspace{1mm}} c @{\hspace{1mm}} l}
      \tr (v^t \tp u^s) & = & \tr (v^t) \; \tr (u^s) \:, \\
        & & \hspace{1cm}
          s=\jnd,1,\ldots,\frac{N_0}{2}-\jnd \:, \;\;\; t=\jnd,1,\ldots \:, \\
      \tr (v^{t+\jnd}) & = & \tr (v^t) \; \tr (v) - \tr (v^{t-\jnd}) \:, \\
      \tr (v) & = & \tr (u^{\frac{N_0}{2}-\jnd}) \; \tr (u) -
                2 \: \tr (u^{\frac{N_0}{2}-1}) \:, \\
      \tr (u^{s+\jnd}) & = & \tr (u^s) \; \tr (u) - \tr (u^{s-\jnd}) \:, \\
        & & \hspace{1cm}
          s=\jnd,1,\ldots,\frac{N_0}{2}-1 \:, \;\;\; t=\jnd,1,\ldots \:, \\
      \tr (u) & = & \al + \de \:.
    \end{array}
  \end{equation}
  \renewcommand{\arraystretch}{1}%

  We get that the~statement of the~lemma is equivalent to
  \begin{equation}
    \hlabel{i:AcentAlDe}
    A_{\rm cent} =
          \spn \left\{ (\al + \de)^n \: : \: n=0,1,2,\ldots \right\} \:.
  \end{equation}

  Let us consider a~linear mapping $\psi : A \rightarrow A \ot A$ defined~as
  \begin{equation}
    \psi = \Kom - \flip \circ \Kom \:,
  \end{equation}
  The~equation~(\ref{i:AcentAlDe}) is equivalent to
  \begin{equation}
    \hlabel{i:kerpsi}
    \ker \psi = \spn \left\{ (\al + \de)^n \: : \: n=0,1,2,\ldots \right\}.
  \end{equation}
  Let $\psiP = \restr{\psi}{A_P}$ for $P=0,1,2,\ldots$.
  The~equation~(\ref{i:kerpsi}) can be replaced by the~following
  \[
    \forall \; P=0,1,2,\ldots \hspace{1cm}
    \ker \psiP =
    \spn \left\{ (\al + \de)^n \: : \: n=0,1,\ldots,P \right\} \:.
  \]
  Note that $\dim \ker \psiP \geq P+1$
  (because $(\al + \de)^n \in \ker \psiP$).
  Thus the~above equality follows from the~inequality
  \begin{equation}
    \hlabel{i:diminequality}
    \dim \image \psiP \geq \dim A_P - (P+1) \:,
  \end{equation}
  which we are going to prove now.

  The~comultiplication $\Kom$ of the~Hopf algebra $A$ is given by
  \begin{equation}
    \hlabel{i:comultiplication}
    \begin{array}{ccc}
      \Kom(\al) = \al \ot \al + \be \ot \ga \:, & &
      \Kom(\de) = \ga \ot \be + \de \ot \de \:, \\
      \Kom(\be) = \al \ot \be + \be \ot \de \:, & &
      \Kom(\ga) = \ga \ot \al + \de \ot \ga \:. \\
    \end{array}
  \end{equation}

  Let us take into account elements of the~form
  \[
    \Kom(\al^k \de^l \be_m) \:, \;\;\;
    k,l \in \N \:, \; m \in \Z \:, \;\; \mbox{such that} \;\;
    k + l + |m| \leq p \:,
  \]
  for certain $p \in \{ 0,1,2,\ldots,P \}$.
  There are two cases
  \begin{enumerate}
    \item $m \neq 0$. Using~(\ref{c:bem}) and~(\ref{i:comultiplication})
          one has
    \begin{displaymath}
      \renewcommand{\arraystretch}{1.4}%
      \begin{array}{c}
        \Kom(\al^k \de^l \be^m) =
            \underline{   c \al^{k+m} \ga^l \ot \al^k \be^{m+l}   } +
            (\mbox{\small elements with at least one $\de$}) + x \:, \\
        \Kom(\al^k \de^l \ga^m) =
            \underline{   d \al^k \ga^{m+l} \ot \al^{k+m} \be^l   } +
            (\mbox{\small elements with at least one $\de$}) + y \:, \\
        \mbox{where} \;\;\; c, d \in \C \:, \;\; c, d \neq 0 \:, \;\;
        x, y \in A_{p-1} \ot A_p + A_p \ot A_{p-1} \:.
      \end{array}
      \renewcommand{\arraystretch}{1}%
    \end{displaymath}

    \item $m=0$, \ $l\geq 1$.
    \begin{displaymath}
      \renewcommand{\arraystretch}{1.4}%
      \begin{array}{c}
        \Kom(\al^k \de^l) =
            \underline{   \al^k \ga^l \ot \al^k \be^l    } +
            (\mbox{\small elements with at least one $\de$}) + x \:, \\
        \mbox{where} \;\;\; x \in A_{p-1} \ot A_p + A_p \ot A_{p-1} \:.
      \end{array}
      \renewcommand{\arraystretch}{1}%
    \end{displaymath}
  \end{enumerate}
  One can easily see that the~underlined elements
  and the~elements one gets acting with~$\flip$ on,
  are linearly independent for
  $k,l \in \N$, \ $m \in \Z$ \ such that
  $k + l + |m| \leq p$ ($m \neq 0$ or $l \geq 1$), $p=0,1,\ldots,P$
  (see the~basis~(\ref{c:elementybazy2})).
  Hence the~elements~$\Psi_p(\al^k \de^l \be_m)$
  are also linearly independent. We have not considered only the~elements
  $\Psi_p(\al^k)$, $k=0,1,\ldots,P$.
  Thus we just proved~(\ref{i:diminequality}) as well as
  the~Lemma~\ref{i:Acent}.
\cnd


\hsubsection{More about irreducible representations}
\hlabel{o:MoreAboutIrrRep}

In~this~subsection we describe the~"diagonal part" of~tensor product
of~any two irreducible representations of~$SL_q(2)$.

Using (\ref{c:xwar}) one gets that
\begin{equation}
  \hlabel{o:basisKs}
  e_{(i)} =
  \sum_{
    \begin{array}{c}
      \scriptstyle i_1, i_2, \ldots, i_{2s} = 1,2 \\
      \scriptstyle \SetPow{ \{k:i_k=2\} }=i
    \end{array}
  }
  q^{ - \SetPow{ \{ \; (m,s) \; : \; m<s, \;\; i_m > i_s \; \} } } \;
  e_{i_1} \ot \ldots \ot e_{i_{2s}} \:,
\end{equation}
$i=0,1,\ldots,2s$, form a~basis of $K^s$,
where $\SetPow{B}$ denotes the~number of elements in a~set~$B$.
Analogous basis elements for $v^t$ are called
\begin{equation}
  \hlabel{o:basisKt}
  e'_{(i)} \:, \;\;\; i=0,1,\ldots,2t \:,
\end{equation}
where $e_k, \; k=1,2$ are replaced with
\begin{equation}
  \hlabel{o:basisKtDef}
  e'_k = \underbrace{ e_k \ot \ldots \ot e_k }_{
         \mbox{\scriptsize $N_0$ \ times}    } \:,  \;\;\;\;\; k=1,2 \:,
\end{equation}
and $q$ is replaced by $q'=q^{N_0^2}$ in the~formula (\ref{o:basisKs}) .

\begin{fakt}
  \hlabel{o:UV_VUfakt}
  The~representations $v^t \tp u^s$ and $u^s \tp v^t$, \
  $t=0,\jnd,1,\ldots$, \linebreak
  $s=0,\jnd,1,\ldots,\frac{N_0}{2}-\jnd$, \
  are equivalent. An~invertible interwiner~$S$ satisfying
  \[
    (S \ot I) (u^s \tp v^t) = (v^t \tp u^s) (S \ot I)
  \]
  is
  (in the~bases consisting of tensor products of~(\ref{o:basisKs})
       and~(\ref{o:basisKt}))
  given by
  \[
    S_{ij,mn} = (q^{N_0})^{2 j t + 2 i s}  \de_{in} \de_{jm} \:,
  \]
  where $\de_{a b} = 1$ for $a=b$ and $0$ otherwise.
  Note that $q^{N_0} = \pm 1$.
\end{fakt}

\dow
  One can easily compute the~rules
  (see~(\ref{c:relations}), cf~Remark~\ref{m:CentrumA}.)
  \begin{equation}
    \hlabel{o:AlBeGaDeRelations}
\renewcommand{\tmp}{@{\hspace{1mm}}}
    \begin{array}{r \tmp c \tmp l r \tmp c \tmp l r \tmp c \tmp l}
      \al^{N_0} \: \de & = & \de \: \al^{N_0} \:, &
      \al^{N_0} \: \be & = & q^{N_0} \: \be \: \al^{N_0} \:, &
      \al^{N_0} \: \ga & = & q^{N_0} \: \ga \: \al^{N_0} \:, \\
      \de^{N_0} \: \al & = & \al \: \de^{N_0} \:, &
      \be \: \de^{N_0}  & = & q^{N_0} \: \de^{N_0} \: \be \:, &
      \ga \: \de^{N_0} & = & q^{N_0} \: \de^{N_0} \: \ga \:, \\
      \al \: \be^{N_0} & = & q^{N_0} \: \be^{N_0} \: \al \:, &
      \be^{N_0} \: \de & = & q^{N_0} \: \de \: \be^{N_0} \:, &
      \be^{N_0} \: \ga & = & \ga \: \be^{N_0} \:, \\
      \al \: \ga^{N_0} & = & q^{N_0} \: \ga^{N_0} \: \al \:, &
      \ga^{N_0} \: \de & = & q^{N_0} \: \de \: \ga^{N_0} \:, &
      \ga^{N_0} \: \be & = & \be \: \ga^{N_0} \:.
    \end{array}
  \end{equation}

  One has
  \begin{equation}
    \hlabel{o:ue12}
    \begin{array}{c}
      u \; e_1 = e_1 \ot \al + e_2 \ot \ga \:, \\
      u \; e_2 = e_1 \ot \be + e_2 \ot \de \:.
    \end{array}
  \end{equation}
  Let $(u^s)_{ij} \in A$ be matrix elements of~$u^s$ given
  in~the~basis~(\ref{o:basisKs}),
  i.e.\  \linebreak
  $u^{\tps 2s} e_{(j)} = \sum_{i=0}^{2s} \; (u^s)_{ij} \; e_{(i)}$.
  Then one has
  \begin{equation}
    \hlabel{o:basisUsum}
    (u^s)_{ij} = \sum_{
      \begin{array}{c}
        \scriptstyle k \in \Z, \;\; l,r \in \N \\
        \scriptstyle l + r = i - j \;\; (\mbox{\scriptsize mod $2$}) \\
        \scriptstyle |k| + l + r = 2s \;\; (\mbox{\scriptsize mod $2$})
      \end{array}
    }
    a_{klr} \al_k \be^l \ga^r \:, \;\;\; a_{klr} \in \C \:.
  \end{equation}
  The~only elements that change the~quantity of $e_2$ in $e_{(j)}$
  are $\be$ and $\ga$ (see~(\ref{o:ue12}))
  (this corresponds to the~first condition in the~above sum).
  If $\al$ "meets" $\de$ they produce $\be \ga$ and~$I$
  (see~\ref{c:relations})
  (this corresponds to the~second condition).

  Similarly, replacing~(\ref{o:basisKs}) by~(\ref{o:basisKt}), one gets
  that $(v^t)_{mn}$ is a~linear combination of
  $\al_{N_0 k'} \be^{N_0 l'} \ga^{N_0 r'}$ with
  $l'+r'=m-n (\mbox{mod $2$})$,
  $|k'|+l'+r'=2t (\mbox{mod $2$})$.

  Using~(\ref{o:basisKs}), (\ref{o:AlBeGaDeRelations})
  and (\ref{o:basisUsum}),
  one can obtain
  \[
    (u^s)_{ij} (v^t)_{mn} =
    (q^{N_0})^{ 2 s (n-m) + 2 t (j-i) } (v^t)_{mn} (u^s)_{ij} \:.
  \]
\cnd

\uwaga The~equivalence of representations $v^t \tp u^s$ and $u^s \tp v^t$
  follows also from ~(\ref{i:tracesUV}) and Proposition~\ref{q:trUtrVfakt}.


\vspace{1ex}
Using Proposition~\ref{o:UV_VUfakt}.,
Corollary~\ref{m:uTdecomposition}.\ and
Corollary~\ref{m:twVrozklad}.\ one gets
\begin{fakt}
  \hlabel{o:VUtUV}
  \vspace{-5mm}
  \begin{eqnarray*}
    &
    (v^t \tp u^s) \tpt (v^{t'} \tp u^{s'}) \approx
    \raisebox{-1.2ex}{$
      \begin{array}{c}
        \scriptstyle {t+t'} \\
        \oplus \\
        \scriptstyle {r=|t-t'|} \\
        \mbox{\scriptsize step $ = 1 $}
      \end{array}
    $}
    \raisebox{-1.2ex}{$
      \begin{array}{c}
        \scriptstyle {s+s'} \\
        \oplus \\
        \scriptstyle {r'=|s-s'|} \\
        \mbox{\scriptsize step $ = 1 $}
      \end{array}
    $}
    v^r \tp \widetilde{u^{r'}} \:,
    & \\
    &
    t, t' = 0,\jnd,1,\ldots \:, \;\;\;
    s, s' = 0,\jnd,1,\ldots,\frac{N_0}{2}-\jnd \:,
    &
  \end{eqnarray*}
  (for $s=s'=0$ we can omit~$\WT$).
\end{fakt}

\begin{fakt}
  \hlabel{o:wildUfact}
  \renewcommand{\arraystretch}{1.4}%
  \[
    \begin{array}{r @{\hspace{1mm}} c @{\hspace{1mm}} l}
      \widetilde{ u^{t N_0 + s} } & \approx &
        \left( v^{t-\jnd} \tp u^{\frac{N_0}{2}-s-1} \right) \oplus
        \left( v^t \tp u^s \right) \:, \\
      \widetilde{ u^{t N_0 + (\frac{N_0}{2}-\jnd) } } & \approx &
        \left( v^t \tp u^{\frac{N_0}{2}-\jnd} \right) \:, \\
      \multicolumn{3}{c}{
          t = \jnd,1,\ldots \:, \;\;\;
          s = 0,\jnd,1,\ldots,\frac{N_0}{2} - 1 \:,
      }
    \end{array}
  \]
  \renewcommand{\arraystretch}{1}%
where the~representation $v^t \tp u^s$ is not
  a~subrepresentation of $u^{t N_0 + s}$ (this is only the~quotient
  representation).
\end{fakt}

\dow
  Using mathematical induction one can easily prove
  the~decomposition of the~thesis
  (cf~Corollary~\ref{m:uTdecomposition}.\
  and Theorem~\ref{m:twUVrozklad}.2.).
  It~remains to prove that the~representation $v^t \tp u^s$ is not
  a~subrepresentation of $u^{t N_0 + s}$.

  Let $B$ be an~algebra with unity~$I$ generated by two elements
  $a$,~$a^{-1}$ such that $a \: a^{-1} = a^{-1} \: a = I$.
  Let $\ABHom$ be a~homomorphism of $A$ into $B$ such that
\renewcommand{\tmp}{\hspace{0.8cm}}
  \[
    \ABHom(\al) = a \:, \tmp \ABHom(\de) = a^{-1} \:, \tmp
    \ABHom(\be) = 0 \:, \tmp \ABHom(\ga) = 0 \:.
  \]
  One can obtain~(cf~(\ref{o:ue12}))
  \renewcommand{\arraystretch}{1.4}
  \[
    \begin{array}{c}
      \ABHom (u) \; e_1 = e_1 \ot a \:, \;\;\;
      \ABHom (u) \; e_2 = e_2 \ot a^{-1} \:, \\
      \ABHom (u^s) \; e_{(i)} = e_{(i)} \ot a^{2s-2i} \:, \;\;\;
      \ABHom (v^s) \; e'_{(i)} = e'_{(i)} \ot a^{(2s-2i) N_0} \:, \\
      \hspace{1cm} s=0,\jnd,1,\ldots, \;\;\; i=0,1,\ldots,2s \:.
    \end{array}
  \]
  \renewcommand{\arraystretch}{1}%
Assume that there exists an~invariant subspace $W$ of
  $K^{t N_0 + s}$ corresponding to $v^t \tp u^s$.
  Using $\ABHom$ one can prove that $e_{(0)}$ belongs to~$W$.
  On~the~other hand (cf~(\ref{o:basisKs}) and~(\ref{o:ue12}))
  \[
    u^r e_{(0)} = \sum_{i=0}^{2r} e_{(i)} \ot \al^{2r-i} \ga^i \:,
  \]
  where \ $r=t N_0 + s$ \ in our case
  (it suffices to compare the~elements multiplying
  $e_1 \ot \ldots e_1 \ot e_2 \ot \ldots e_2$).
  The~coefficients $\al^{2r-i} \ga^i$ ($i=0,1,\ldots,r$) are linearly
  independent~(see~Proposition~\ref{c:baza}.).
  Thus $W$ is not an~invariant subspace.
\cnd

\uwaga A~related fact at the~level of~universal enveloping algebras is
given in Proposition~9.2~of~\cite{b:Lusztig}.

\vspace{1ex}
One can prove
\[
   S_M ( \si_k - \id ) = 0  \;\;\;\;
   \mbox{for $k=1,\ldots, M-1$, $M \in \N$ }
\]
in a~similar way as Proposition~\ref{c:si-id}.
Let $i_1,i_2,\ldots,i_M = 1,2$ be such that $ \SetPow{ \{k:i_k=2\} } = m $.
Using (\ref{c:si()}) one gets
\[
 S_M \; e_{i_1} \ot \ldots \ot e_{i_2}  =
  q^{ - \SetPow{ \{ \; (r,t) \; : \; r<t, \;\; i_r > i_t \; \} } } \;
  S_M
  e_1 \ot \ldots \ot e_1 \ot
  \underbrace{e_2 \ot \ldots \ot e_2}_{
     \mbox{\scriptsize $m$ \ times}
  } \:.
\]
On the~other hand (cf~Corollary~\ref{c:imageSN}.\
and the~proof of Proposition~\ref{c:dimImSNg}.)
\[
  S_M
  e_1 \ot \ldots \ot e_1 \ot
  \underbrace{e_2 \ot \ldots \ot e_2}_{
     \mbox{\scriptsize $m$ \ times}
  } =
  \fact_{\frac{1}{q}} (M-m) \; \fact_{\frac{1}{q}} (m) \; e_{(m)} \:.
\]
Thus $\image S_M = \spn \: \{ e_{(m)} \: : \: M-N_0 < m < N_0 \}$.
Therefore $\image S_{N_0 + 2s}$ is a~carrier vector space of
$u^{\frac{N_0}{2}-s-1}$ in~Proposition~\ref{o:wildUfact}.\
for $t=\jnd$, \ $0 \leq s \leq \frac{N_0}{2} - 1$.
Moreover, $S_M = 0$ for $M \geq 2 N_0-1$.


\hsubsection{Haar measure}

\begin{twierdzenie}
  The~quantum group $SL_q(2)$ does not have the~Haar functional.
\end{twierdzenie}

\dow
  Assume that the~Haar functional $h$ does exist.
  Let us take into consideration the~representation
  $\widetilde{u^{t N_0 + s}}$
  of Proposition~\ref{o:wildUfact}.\ for $t=\jnd$, \ $s=\frac{N_0}{2}-1$.
  One has an~explicit form
  \[
    \widetilde{u^{N_0-1}} \approx I \oplus ( v \tp u^{\frac{N_0}{2}-1} ) \:,
  \]
  where $I$ is a~subrepresentation of~$u^{N_0-1}$.
  Applying $h$ to~(\ref{d:HaarRule}) one obtains \linebreak
  $(h \ot h) \Kom = h$,
  hence $P=(\id \ot h) u^{N_0-1}$ is a~projection.
  But
  \begin{eqnarray*}
    \tr P & = & h \tr u^{N_0-1}
            =   h \tr \widetilde{ u^{N_0-1} }  \\
          & = & \tr (\id \ot h)
              \left[
                I \oplus \left( v \tp u^{\frac{N_0}{2} - 1} \right)
              \right]    \\
          & = & \tr (1 \oplus 0 \oplus \ldots \oplus 0)
            = 1 \:,
  \end{eqnarray*}
  and $P \approx 1 \oplus 0 \oplus \ldots \oplus 0$.
  Applying~(\ref{d:HaarRule}) to $u^{N_0-1}$, one obtains
  $
    P u^{N_0-1} = u^{N_0-1} P = P \: I
  $
  and
  \[
    u^{N_0 - 1}
    \approx I \oplus \left( v \tp u^{\frac{N_0}{2} - 1} \right)
  \]
  in contradiction with Proposition~\ref{o:wildUfact}.
\cnd


\hsubsection{Enveloping algebra}
\hlabel{e:EnvelopingAlgebra}

In~this~subsection we describe representations of~$\Uqsl$ corresponding
to the~irreducible representation of~$SL_q(2)$.

Let us recall (cf~\cite{b:Jimbo}, \cite{b:RA})
that quantum universal enveloping algebra $U=\Uqsl$ is the~algebra
with identity~$I$ generated by $\Kp$, $\Km$, $\Jp$, $\Jm$, satisfying
\renewcommand{\arraystretch}{1.4}%
\[
  \begin{array}{c}
    \Kp \: \Km = \Km \: \Kp = I \:, \\
    \Kp \: \Jpm \: \Km = q^{\pm 1} \: \Jpm \:, \\
    \left[ \Jp, \Jm \right] = \frac{(\Kp)^2 - (\Km)^2}{q-q^{-1}}
  \end{array}
\]
\renewcommand{\arraystretch}{1}%
with Hopf algebra structure given by
\renewcommand{\arraystretch}{1.4}%
\[
  \begin{array}{r @{\hspace{1mm}} c @{\hspace{1mm}} l}
    \Kom \: \Kpm & = & \Kpm \ot \Kpm \:, \\
    \Kom \: \Jpm & = & \Kp \ot \Jpm + \Jpm \ot \Km \:.
  \end{array}
\]
\renewcommand{\arraystretch}{1}%
There exists (cf~\cite{b:RTF}, \cite{b:Lyakhovskaya}) unique bilinear
pairing
$U \times A \ni (l,a) \rightarrow \pair{ l }{ a } \in \C$ satisfying
\renewcommand{\arraystretch}{1.4}%
\[
  \begin{array}{r @{\hspace{1mm}} c @{\hspace{1mm}} l}
    \pair{ l_1 l_2 }{ a } & = & \pair{ l_1 \ot l_2 }{ \Kom a } \:, \\
    \pair{ l }{ a_1 a_2 } & = & \pair{ \Kom l }{ a_1 \ot a_2 } \:, \\
    \multicolumn{3}{c}{
      l, l_1, l_2 \in U \:, \;\;\; a, a_1, a_2 \in A \:,
    }
  \end{array}
\]
\renewcommand{\arraystretch}{1}%
\renewcommand{\arraystretch}{1.4}%
\[
  \begin{array}{r @{\hspace{1mm}} c @{\hspace{1mm}} l @{\hspace{8mm}}
                r @{\hspace{1mm}} c @{\hspace{1mm}} l}
    \pair{ \Kp }{ \al } & = & q^{\jnd} \:, &
    \pair{ \Kp }{ \de } & = & q^{-\jnd} \:, \\
    \pair{ \Km }{ \al } & = & q^{-\jnd} \:, &
    \pair{ \Km }{ \de } & = & q^{\jnd} \:, \\
    \pair{ \Kp }{ I } & = & 1 \:, &
    \pair{ \Km }{ I } & = & 1 \:, \\
    \pair{ \Jm }{ \ga } & = & q^{-\jnd} \:, &
    \pair{ \Jp }{ \be } & = & q^{\jnd} \:,
  \end{array}
\]
\renewcommand{\arraystretch}{1}%
$\Kp, \Km, \Jp, \Jm$ vanish at $\al, \be, \ga, \de, I$ in all other cases
(nonzero $l^{\pm}_{ij}$ of~\cite{b:RTF} are given by
\renewcommand{\arraystretch}{1.4}%
\[
  \begin{array}{r @{\hspace{1mm}} c @{\hspace{1mm}} l @{\hspace{8mm}}
                r @{\hspace{1mm}} c @{\hspace{1mm}} l}
    l^+_{11}(a) = l^-_{22}(a) & = & \pair{ \Kp }{ a } \:, &
    l^+_{12}(a) & = & (q-q^{-1}) \: \pair{ \Jm }{ a } \:, \\
    l^+_{22}(a) = l^-_{11}(a) & = & \pair{ \Km }{ a } \:, &
    l^-_{21}(a) & = & - (q-q^{-1}) \: \pair{ \Jp }{ a } \:,
  \end{array}
\]
\renewcommand{\arraystretch}{1}%
$a \in A$. $X_{\pm}$, $K$ of~\cite{b:Lyakhovskaya} correspond to
$q^{\mp \jnd} \Jpm$, $\Kp$ ).
We shall write $l(a)$ instead of $\pair{l}{a}$.

For any representation $w$ of~the~quantum group we introduce
a~unital representation $\Pi_w$
of the~algebra $U$ by
\[
  \left[ \Pi_w (l) \right]_{ij} = l( w_{ij} ) =
  \left[ (\id \ot l) w \right]_{ij}   \:,
  \;\;\; i,j=1,2,\ldots,\dim w \:.
\]
We are going to describe $\Pi_w$ for any irreducible representation $w$
of~$SL_q(2)$.
Let us fix $s=0,\jnd,1,\ldots,\frac{N_0}{2}-\jnd$. One can obtain
\renewcommand{\arraystretch}{1.4}%
\begin{eqnarray}
\hlabel{e:komK}
  & &
  \Kom^{(2s)} \Kpm =
     \underbrace{\Kpm \ot \ldots \ot \Kpm
                }_{\mbox{\scriptsize $2s$ \ times}} \:, \\
\hlabel{e:komX}
  & &
  \Kom^{(2s)} \Jpm = \sum_{m=1}^{2s}
    \underbrace{\Kp \ot \ldots \ot \Kp
               }_{\mbox{\scriptsize $m-1$ \ times}}
                \ot \Jpm \ot
    \underbrace{\Km \ot \ldots \ot \Km
               }_{\mbox{\scriptsize $2s-m$ \ times}} \:.
\end{eqnarray}
\renewcommand{\arraystretch}{1}%

Using~(\ref{e:komK}) one can show that
\begin{equation}
  \hlabel{e:Kpm}
  (\id \ot \Kpm) \: u^s \; e_{(j)} = q^{\pm (s-j)} \; e_{(j)} \:, \;\;\;\;
  j=0,1,\ldots,2s \:.
\end{equation}

Using~(\ref{e:komX}) one has
\begin{eqnarray}
\hlabel{e:bigU}
    (\id \ot \Jm) \: u^{\tps 2s} & = & \sum_{m=1}^{2s}
  \underbrace{
    \left( \begin{array}{cc}
       q^{\jnd} & 0        \\
       0        & q^{-\jnd}
    \end{array} \right)
    \ot \ldots \ot
    \left( \begin{array}{cc}
       q^{\jnd} & 0        \\
       0        & q^{-\jnd}
    \end{array} \right)
  }_{\mbox{\scriptsize $m-1$ \ times}}
    \ot
  \\ & & 
    \left( \begin{array}{cc}
       0         & 0 \\
       q^{-\jnd} & 0
    \end{array} \right)
    \ot
  \underbrace{
    \left( \begin{array}{cc}
       q^{-\jnd} & 0        \\
       0        & q^{\jnd}
    \end{array} \right)
    \ot \ldots \ot
    \left( \begin{array}{cc}
       q^{-\jnd} & 0        \\
       0        & q^{\jnd}
    \end{array} \right)
  }_{\mbox{\scriptsize $2s-m$ \ times}} \vspace{1cm} \:.
  \nonumber
\end{eqnarray}
On the~other hand
\begin{equation}
\hlabel{e:uX}
  (\id \ot \Jm) \: u^s \; e_{(j)}
  = \sum_{m=0}^{2s}
      \Jm( u^s_{mj} ) \: e_{(m)} \:.
\end{equation}
Thus $\Jm(u^s_{mj})$ equals to the~coefficient multiplying
$e_1 \ot \ldots \ot e_1 \ot e_2 \ot \ldots \ot e_2$.
Considering~(\ref{e:bigU}), that~coefficient can
be nonzero only for $m=j+1$, $j<2s$ (cf~(\ref{o:basisKs})).
In that case it is obtained from the~following
explicitly written part of~$e_{(j)}$
\[
  e_{(j)} = \sum_{r=0}^{m-1} q^{-r}
  \underbrace{e_1 \ot \ldots \ot e_1
             }_{\mbox{\scriptsize $2s-m$ \ times}}
  \ot
  \underbrace{  \overbrace{e_2 \ot \ldots \ot e_2
                          }^{\mbox{\scriptsize $r$ \ times}}
                 \ot
                 e_1 \ot e_2 \ot \ldots \ot e_2
             }_{\mbox{\scriptsize $m$ \ times}}
  + \ldots
\]
Doing some computations,
\begin{eqnarray*}
  \Jm( u^s_{j+1,j} )  & = &
  q^{s-1} \frac{1-q^{-2(j+1)}}{1-q^{-2}} \:,
\end{eqnarray*}
$j < 2s$.
Doing similar computations for $\Jp$ and~using~(\ref{e:Kpm}) one obtains
\begin{eqnarray*}
    \Pi_{u^s}(\Kpm) \; e_{(j)} & = & q^{\pm (s-j)} \; e_{(j)} \:, \\
    \Pi_{u^s}(\Jm) \; e_{(j)} & = & q^{s-j-1} [j+1] \; e_{(j+1)} \:, \\
    \Pi_{u^s}(\Jp) \; e_{(j)} & = & q^{j-s} [2s-j+1] \; e_{(j-1)} \:,
\end{eqnarray*}
$j=0,1,\ldots,2s$, where $[l]=\frac{q^l-q^{-l}}{q-q^{-1}}$,
$e_{(2s+1)}=e_{(-1)}=0$. Using~(\ref{e:komK}) and~(\ref{e:komX}), one has
\begin{eqnarray*}
  (\id \ot \Kpm) \: v \; e'_1 & = &
           q^{\pm \frac{N_0}{2}} \; e'_1 \:, \\
  (\id \ot \Kpm) \: v \; e'_2 & = &
           q^{\mp \frac{N_0}{2}} \; e'_2 \:, \\
  (\id \ot \Jpm) \: v \; e'_i & = & 0 \:, \;\;\;\; i=1,2 \:.
\end{eqnarray*}
Therefore~(cf~(\ref{o:basisKs})---(\ref{o:basisKtDef}))
\begin{eqnarray*}
    \Pi_{v^t}(\Kpm) \; e'_{(i)} & = & q^{\pm N_0 (t-i)} \; e'_{(i)} \:, \\
    \Pi_{v^t}(\Jpm) \; e'_{(i)} & = & 0 \:, \;\;\;\; i=0,1,\ldots,2t \:,
\end{eqnarray*}
$t=0,\jnd,1,\ldots$. Thus for $w=v^t \tp u^s$ we have
\renewcommand{\arraystretch}{1.4}%
\[
  \begin{array}{r @{\hspace{1mm}} c @{\hspace{1mm}} l}
    \Pi_{w}(\Kpm) \; e'_{(i)} \ot e_{(j)} & = &
       q^{\pm [N_0 (t-i)+s-j] } \; e'_{(i)} \ot e_{(j)} \:, \\
    \Pi_{w}(\Jm) \; e'_{(i)} \ot e_{(j)} & = &
       q^{N_0 (t-i)+s-j-1} [j+1]  \; e'_{(i)} \ot e_{(j+1)} \:, \\
    \Pi_{w}(\Jp) \; e'_{(i)} \ot e_{(j)} & = &
       q^{N_0 (t-i)+j-s} [2s-j+1] \; e'_{(i)} \ot e_{(j-1)} \:, \\
    \multicolumn{3}{c}{
      i=0,1,\ldots,2t \:, \;\;\;   j=0,1,\ldots,2s \:, \;\;\;
      t=0,\jnd,1,\ldots \:, \;\;\;
      s=0,\jnd,1,\ldots,\frac{N_0}{2}-\jnd \:.
    }
  \end{array}
\]
\renewcommand{\arraystretch}{1}%
Hence
$\Pi_w \approx \pi_{-t} \oplus \pi_{-t+1} \oplus \ldots \oplus \pi_{t}$,
where $\pi_r$, $r \in \Z \slash 2$, is the~\mbox{$(2s+1)$-}\-dimensional
representation of~$U$ described by~(1) in~\cite{b:RA}, with
$\omega=q^{N_0 r} \in \{ 1,-1,i,-i \}$ (for $s=\frac{N_0}{2}-\jnd$
we take~(2)
for even roots or~(4) for odd roots instead of~(1),
$\mu=2(s+N_0 r)$, $\al=\be=0$).


\hsubsection{Enveloping algebra according to Lusztig}
\hlabel{l:EnvelopingAlgAcor2Lusz}

Let us introduce the generators of \cite{b:Lusztig} as follows:
\begin{equation}
  \hlabel{l:lgen}
  E=q^{\jnd} \Kp \Jp \:, \;\;\;
  F=q^{-\jnd} \Jm \Km \:, \;\;\;
  K=\left( \Kp \right)^2 \:.
\end{equation}
One can compute:
\renewcommand{\arraystretch}{1.4}%
\[
 \begin{array}{c}
   K K^{-1} = K^{-1} K = 1 \:, \\
   K E K^{-1} = q^2 E \:, \;\;\; K F K^{-1} = q^{-2} F \:, \\
   EF - FE = \frac{K-K^{-1}}{q-q{-1}} \:,
 \end{array}
\]
\renewcommand{\arraystretch}{1}%
\begin{eqnarray*}
   \Kom E & = & K \ot E + E \ot I \:, \\
   \Kom F & = & F \ot K^{-1} + I \ot F \:, \\
   \Kom K & = & K \ot K \:. \\
\end{eqnarray*}

Let $\UL$ be the~Hopf algebra over $\C(q)$ generated by $E$, $F$,
$K^{\pm 1}$ satisfying the~above relations,
where $q$ is an~indeterminate ($v$ in \cite{b:Lusztig}) commuting with every
element of~$\UL$.

We also have a~bilinear pairing $\UL \times A \rightarrow \C(q^{\jnd})$~:
\[
 \begin{array}{c@{\hspace{10mm}}c}
   \pair{K}{\al} = q \:, &
   \pair{K}{\de} = q^{-1} \:, \\
   \pair{K^{-1}}{\al} = q^{-1} \:, &
   \pair{K^{-1}}{\de} = q \:, \\
   \pair{E}{\be} = q^{\frac{3}{2}} \:,  &
   \pair{F}{\ga} = q^{-\frac{3}{2}} \:, \\
   \pair{K}{I} = \pair{K^{-1}}{I} = 1
 \end{array}
\]
and zero for the~rest of combinations.
According to~\cite{b:Lusztig} $\UA$~is~$\C[q,q^{-1}]$ algebra generated by:
\[
   K \:, \;\; K^{-1} \:, \;\; E \:, \;\; F \:, \;\;\;
   E^{(M)}=\frac{E^M}{[M]!} \;, \;\;\; F^{(M)}=\frac{F^M}{[M]!} \;,
\]
where $M=2,3,\ldots$ and
$[M]! = \prod_{j=1}^M \frac{q^j-q^{-j}}{q-q^{-1}} =
        q^{-\frac{M^2}{2}+\frac{M}{2}} \: \fact_q(M) \in \C[q,q^{-1}]$.
We think about~$A$ also as $\C[q,q^{-1}]$ algebra.

Using the~mathematical induction w.r.t. $n,l,m$ one can derive the~formula
(a~corrected version of~(13) of~\cite{b:Lyakhovskaya})
for a~bilinear pairing $\UA \times A \rightarrow \C[q^{\jnd},q^{-\jnd}]$~:
\renewcommand{\arraystretch}{1.4}%
\begin{eqnarray}
  \nonumber
  \lefteqn{ \pair{ E^k K^t F^i }{ \al_n \be^l \ga^m } } \\
    & = & \!\!\! \left\{
    \begin{array}{l@{\hspace{3mm}}l}
      [k]! \; [i]! \;
         q^{ \frac{3}{2}(k-i)-t(k+i)+n(t+k) } \de_{kl} \de_{im}
               & , \; n \le 0 \\
      { [k]! } \; [i]! \;
         q^{ \frac{3}{2}(k-i)-t(k+i)+n(t+k)
         - (k-l)^2 } & \\
      \hspace{3mm} \cdot \;\;
      \frac{ \fact_{\frac{1}{q}}(n) }{
             \fact_\frac{1}{q}(k-l) \fact_\frac{1}{q}(n-k+l) } & \\
      \hspace{3mm} \cdot \;\;
      ( \de_{k-l,0} + \de_{k-l,1} + \ldots + \de_{k-l,n} )  \; \de_{k-l,i-m}
      & , \; n > 0
    \end{array}
    \right.
  \hlabel{l:PairForm}
\end{eqnarray}
\renewcommand{\arraystretch}{1}%

In the~same way as in section~\ref{e:EnvelopingAlgebra} we get:
\begin{eqnarray*}
    \Pi_{u^s}(K^{\pm 1}) \; e_{(j)} & = & q^{\pm 2(s-j)} \; e_{(j)} \:, \\
    \Pi_{u^s}(E) \; e_{(j)} & = &
                             q^{\frac{3}{2}} [2s-j+1] \; e_{(j-1)} \:, \\
    \Pi_{u^s}(F) \; e_{(j)} & = & q^{-\frac{3}{2}} [j+1] \; e_{(j+1)} \:,
\end{eqnarray*}
$j=0,1,\ldots,2s$, $s=0,\jnd,1,\ldots,\frac{M}{2}-\jnd$,
where $e_{(2s+1)}=e_{(-1)}=0$.
Matrix
$v = \left(
  \matrix{ \al^M & \be^M \cr \ga^M & \de^M \cr }
     \right)  $
(see~(\ref{m:vdef})) is not a~representation at the~moment, since $q$ is
an~indeterminate.
Using~(\ref{l:PairForm}) we get:
\renewcommand{\arraystretch}{2.2}%
\[
 \begin{array}{c}
   \pair{E}{v} = \pair{F}{v} = \left( \matrix{ 0 & 0 \cr 0 & 0 \cr }
                               \right) \:, \\
   \pair{K^{\pm 1}}{v} =
       \left( \matrix{ q^{\pm M} & 0 \cr 0 & q^{\mp M} \cr }
                 \right) \:. \\
 \end{array}
\]
\renewcommand{\arraystretch}{1}%
Hence
\begin{eqnarray*}
    \Pi_{v^t}(K^{\pm 1}) \; e'_{(i)} & = & q^{\pm 2 M (t-i)} \; e'_{(i)}
    \:, \\
    \Pi_{v^t}(E) \; e'_{(i)} & = & 0 \:, \\
    \Pi_{v^t}(F) \; e'_{(i)} & = & 0 \:, \\
\end{eqnarray*}
$t=0,\jnd,1,\ldots$, $i=0,1,\ldots,2t$.
Thus for $w=v^t \tp u^s$ we have
\renewcommand{\arraystretch}{1.4}%
\[
  \begin{array}{r @{\hspace{1mm}} c @{\hspace{1mm}} l}
    \Pi_{w}(K^{\pm 1}) \; e'_{(i)} \ot e_{(j)}
              & = & \pair{K^{\pm 1}}{v^t} \pair{K^{\pm 1}}{u^s} \;
                    e'_{(i)} \ot e_{(j)} \\
              & = & q^{\pm 2((s-j)+M(t-i))} \; e'_{(i)} \ot e_{(j)} \:, \\
    \Pi_{w}(E) \; e'_{(i)} \ot e_{(j)}
              & = & \pair{K}{v^t} \pair{E}{u^s} \; e'_{(i)} \ot e_{(j)} \\
              & = & q^{\frac{3}{2}+2M(t-i)} [2s-j+1] \;
                    e'_{(i)} \ot e_{(j-1)} \:, \\
    \Pi_{w}(F) \; e'_{(i)} \ot e_{(j)}
              & = & \pair{I}{v^t} \pair{F}{u^s} \; e'_{(i)} \ot e_{(j)} \\
              & = & q^{-\frac{3}{2}} [j+1]  \;
                    e'_{(i)} \ot e_{(j+1)} \:, \\
    \multicolumn{3}{c}{
      i=0,1,\ldots,2t \:, \;\;\;   j=0,1,\ldots,2s \:, \;\;\;
      t=0,\jnd,1,\ldots \:, \;\;\;
      s=0,\jnd,1,\ldots,\frac{M}{2}-\jnd \:.
    }
  \end{array}
\]
\renewcommand{\arraystretch}{1}%
Using mathematical induction we can prove that
$\pair{E^k}{u^{\tps 2s}}=0$ for $2s<k$. In particular we have
\[
  \pair{E^{(M)}}{u^s}=0 \;\;\; \mbox{for} \; s < \frac{M}{2} \:.
\]
In analogous way we get
\[
  \pair{F^{(M)}}{u^s}=0 \;\;\; \mbox{for} \; s < \frac{M}{2} \:.
\]
Using formula
$(a+b)^M=\sum_{r=0}^M \frac{\fact_q(M)}{\fact_q(r)\fact_q(M-r)} a^r b^{M-r} $,
where $ba = q^2 ab$, we get
\begin{equation}
  \hlabel{l:ENkomn}
  \Kom (E^{(M)}) = \left(
       \sum_{r=1}^{M-1} \frac{q^{r(M-r)}}{[r]![M-r]!}
       E^r K^{M-r} \ot E^{M-r}    \right)
       + K^M \ot E^{(M)} + E^{(M)} \ot I \:.
\end{equation}
In virtue of~(\ref{l:PairForm})
\renewcommand{\arraystretch}{1.4}%
\[
 \begin{array}{c}
    \pair{E^{(M)}}{v} = \left(
          \matrix{ 0 & q^{\frac{3}{2}M} \cr 0 & 0 \cr }  \right)
    \:, \hspace{5mm}
    \pair{F^{(M)}}{v} = \left(
          \matrix{ 0 & 0 \cr q^{-\frac{3}{2}M} & 0 \cr }  \right) \:.
 \end{array}
\]
\renewcommand{\arraystretch}{1}%
Using the~formula~$\pair{E^r K^{M-r}}{v}=0$
for~$r=1,\ldots,M-1$, (\ref{l:ENkomn}) and doing some computations, we get
\begin{eqnarray}
  \nonumber
  \pair{E^{(M)}}{v^{\tps2t}}  & = & \sum_{m=1}^{2t}
  \underbrace{
    \left( \begin{array}{cc}
       q^{M^2} & 0        \\
       0        & q^{-M^2}
    \end{array} \right)
    \ot \ldots \ot
    \left( \begin{array}{cc}
       q^{M^2} & 0        \\
       0        & q^{-M^2}
    \end{array} \right)
  }_{\mbox{\scriptsize $m-1$ \ times}}
    \ot
  \\ & & 
    \left( \begin{array}{cc}
       0 & q^{\frac{3}{2}M} \\
       0 & 0
    \end{array} \right)
    \ot
  \underbrace{
    \left( \begin{array}{cc}
       1 & 0 \\
       0 & 1
    \end{array} \right)
    \ot \ldots \ot
    \left( \begin{array}{cc}
       1 & 0 \\
       0 & 1
    \end{array} \right)
  }_{\mbox{\scriptsize $2t-m$ \ times}} \vspace{1cm} \:.
\end{eqnarray}
Let us now specialize $q$ to be a~root of unity of~degree $N \geq 3$
(cf \cite{b:Lusztig}). Let $M=N_0$.
Doing similar comparison of coefficients
as in~(\ref{e:bigU}) and~(\ref{e:uX}), we get
\begin{eqnarray*}
  \Pi_{v^t}(E^{(N_0)}) \; e'_{(i)} & = & q^{\frac{3}{2} N_0 - i N_0^2} \;
    (2t-i+1) \;
    e'_{(i-1)}
  \:, \;\;\;\;  i=0,1,\ldots,2t \:, \\
  \Pi_{v^t}(F^{(N_0)}) \; e'_{(i)} & = & q^{-\frac{3}{2} N_0 + i N_0^2} \;
    (i+1) \;
    e'_{(i+1)}
  \:, \;\;\;\;  i=0,1,\ldots,2t \:, \\
\end{eqnarray*}
where $t=0,\jnd,1,\ldots$.
Thus for $w=v^t \tp u^s$ we get~(cf~(\ref{l:ENkomn})):
\renewcommand{\arraystretch}{1.4}%
\[
  \begin{array}{r @{\hspace{1mm}} c @{\hspace{1mm}} l}
    \Pi_{w}(E^{(N_0)}) \; e'_{(i)} \ot e_{(j)}
              & = & \pair{E^{(N_0)}}{v^t} \pair{I}{u^s} \;
                    e'_{(i)} \ot e_{(j)} \\
              & = & q^{\frac{3}{2}N_0 - i N_0^2} (2t-i+1) \;
                    e'_{(i-1)} \ot e_{(j)} \:, \\
    \Pi_{w}(F^{(N_0)}) \; e'_{(i)} \ot e_{(j)}
              & = & \pair{F^{(N_0)}}{v^t} \pair{K^{-N_0}}{u^s} \;
                    e'_{(i)} \ot e_{(j)} \\
              & = & q^{-\frac{3}{2}N_0 + i N_0^2 - 2 s N_0 } (i+1)  \;
                    e'_{(i+1)} \ot e_{(j)} \:, \\
    \multicolumn{3}{c}{
      i=0,1,\ldots,2t \:, \;\;\;   j=0,1,\ldots,2s \:, \;\;\;
      t=0,\jnd,1,\ldots \:, \;\;\;
      s=0,\jnd,1,\ldots,\frac{N_0}{2}-\jnd \:.
    }
  \end{array}
\]
\renewcommand{\arraystretch}{1}%

Comparing with representations~of~\cite{b:Lusztig} (in the~case of $N$ odd
as~in~\cite{b:Lusztig})
we infer that $w$~is
isomorphic to the~representation~$L_q(N 2 s + 2 t)$.
Irreducibility of the~obtained representations is related to
the~Lemma~6.1 of~\cite{b:CL}.


\appendix

\hasection{Basic concepts}
\hlabel{d:basicqg}

Here we recall the~basic facts concerning quantum groups and their
representations~(see~\cite{b:CompactMatrix}).

Let $(A,\Kom)$ be a~Hopf algebra.
We set $\Kom^{(2)} = \Kom$,
\[
  \Kom^{(n)} = (\Kom \ot \id \ot \ldots \ot \id) \Kom^{(n-1)} \:,
  \;\;\; n=3,4,\ldots \:.
\]

Let $K$ be a~finite dimensional vector space over~$\C$ and
$e_1, \ldots, e_d \in K$ its basis. Then
$B(K) \ot A \cong B(\C^d) \ot A \cong M_d(A)$.

One can define a~linear mapping
\[
  \ti : {\rm M}_N(A) \times {\rm M}_N(A)
  \longrightarrow {\rm M}_N(A \ot A)
\]
by
\[
  (v \ti w)_{ij} = \sum_k v_{ik} \ot w_{kj} \:, \;\;\; i,j =1,2,\ldots,N \:.
\]

An~element $v \in B(K) \ot A$ is called
{\em a~representation (in the~carrier vector space~$K$)
of a~quantum group corresponding to a~Hopf algebra}
$(A,\Kom,\kappa,e)$, iff
\begin{eqnarray}
\hlabel{d:DefRep1}
  & ( \id \ot \Kom ) \: v = v \ti v \:, \\
\hlabel{d:DefRep2}
  & ( \id \ot e ) \: v = \id \:.
\end{eqnarray}
One can use the~shortcut {\em representation} having in mind the~above
definition.

\vspace{1.5ex}
Let $v$ and $w$ be representations, $K_v$ and $K_w$ be
its~carrier vector spaces.

An~operator $S \in B(K_v,K_w)$ such that
\[
  (S \ot I) \; v = w \; (S \ot I)
\]
is called {\em a~morphism (intertwiner)} of representations~$v$ and~$w$.

A~set of morphisms intertwining $v$ with~$w$ is denoted by~${\rm Mor}(v,w)$.

Representations $v$ and~$w$ are called {\em equivalent} iff
$Mor(v,w)$ contains an~invertible element. Then we write $v \approx w$.

\vspace{1.5ex}
One can define a~linear mapping
\[
  \tp : {\rm M}_N(A) \times {\rm M}_{N'}(A) \longrightarrow {\rm M}_{N N'}(A)
\]
by
\[
  (v \tp w) _{ik,jl} = v_{ij} w_{kl} \:, \;\;\;
  i,j=1,2,\ldots,N \:, \;\;\; k,l=1,2,\ldots,N' \:,
\]
where $v \in {\rm M}_N(A)$, \ $w \in {\rm M}_{N'}(A)$.

One can check that if $v$ and~$w$ are representations,
then $v \tp w$ is also a~representation.
We denote $v^{\tps k} = v \tp \ldots \tp v$ ($k$ times).

\begin{definicja}
  \hlabel{d:HaarDefinition}
  A~functional $\Haar \in A'$ is called
  a~Haar functional of a~quantum group $(A,\Kom)$ if
  $\Haar(I) = 1$ and
  \begin{equation}
  \hlabel{d:HaarRule}
    \forall \; a \in A \hspace{1cm}
    (\Haar \ot \id) \Kom a = (\id \ot \Haar) \Kom a = \Haar(a) \; I \:.
  \end{equation}
\end{definicja}
One has $\Haar(u) = 0$ for any irreducible representation~$u$
different from~$I$.


\hasection{Quotient representations}
\hlabel{q:qkor}

Here we investigate the~quotient representations and the~operation~$\WT$.

Let $u \in B(K) \ot A$ be given~by
\begin{equation}
  \hlabel{q:umr}
  u = \sum_{r=1}^R \; m_r \ot u_r \:,
\end{equation}
where $K=\C^N$ and
$m_1, m_2, \ldots, m_R \in B(K)$ are linearly independent as well as
$u_1, u_2, \ldots, u_R \in A$.
One can introduce a~linear mapping
$\hat{u} : K \rightarrow K \ot A$ given~by
\begin{equation}
  \hlabel{q:u'mr}
  \forall \; x \in K \hspace{1.5em}
                \hat{u} \, x = \sum_{r=1}^R \; m_r \, x \ot u_r \:.
\end{equation}
This is the~formula~(2.7) in~\cite{b:CompactMatrix}.
$\hat{u}$ corresponds to $u$, because $B(K,K \ot A)$ and $B(K) \ot A$
are canonically isomorphic.

\begin{fakt}
\hlabel{q:korepfakt}
Let $(A,\Kom)$ be a~quantum group with a~counit~$e$ and
$u \in M_N(A)$.
Then the~following three statements are equivalent:
\begin{enumerate}
  \item $u$ is a~representation of~$(A,\Kom)$.
  \item For each $x \in K$
     \begin{tabular}[t]{ll}
       i)  & $ (\id \ot \Kom) \hat{u} \, x =
                        (\hat{u} \ot \id) \hat{u} \, x $, \\
       ii) & $ (\id \ot e) \hat{u} \, x = x$.
     \end{tabular}
\end{enumerate}
\end{fakt}

\dow
The~equivalence of (\ref{d:DefRep1}) and~i) is proved
in~\cite{b:CompactMatrix}. The~equivalence of (\ref{d:DefRep2}) and~ii)
can be proved in a~similar way.
\cnd

Let $u$ be a~representation and $K$ be its carrier space.
We say (as in~\cite{b:CompactMatrix}) that $L$ is $u$-invariant subspace
if and only if
\[
  \widehat{u}(L) \;\; \Subrep \;\; L \ot A \:.
\]
Then the~element $\restr{u}{L} \in B(L) \ot A$ corresponding to
the~restriction $\restr{\widehat{u}}{L} : L \rightarrow L \ot A$
is a~representation acting on~$L$.

The~embedding $L \rightarrow K$ intertwines $\restr{u}{L}$ with~$u$,
$\restr{u}{L}$ is called a~subrepresentation of~$u$
(cf~\cite{b:CompactMatrix}).

\begin{definicja}
\hlabel{q:quotrepdef}
Let $u$ be a~representation, $K$ its carrier vector space,
$L \Subrep K$ $u$-invariant space and let $v = \restr{u}{L}$.
Let $[x] \in \quot{K}{L}$ be an~element of the~quotient space and
$x \in K$ a~representative of the~class.
The~representation~$w$ given~by
\begin{equation}
  \hlabel{q:quotrep}
  \forall \; x \in K \hspace{1.5em}
       \hat{w} \, [x] = \sum_{r=1}^R \; [m_r \, x] \ot u_r \:,
\end{equation}
is called {\em a~quotient representation} and is denoted by $\quot{u}{v}$ .
\end{definicja}

Doing some easy computations one can prove that
the~quotient representation~$w$ is uniquely determined and that
it is in fact a~representation (use Proposition~\ref{q:korepfakt}).

\vspace{5ex}
Let us define a~mapping
\begin{equation}
  \hlabel{q:tilde}
  \WT  \; : \; (\mbox{\small representations}) \longrightarrow
   \left(
     \parbox{3cm}{\centering
          \small completely reducible \\ representations
     }
   \right)
\end{equation}
as follows
\begin{enumerate}
  \item If $u$ is an~irreducible representation, then
      $ \widetilde{u} = u$.
  \item If $v$ is a~subrepresentation of~$u$, then we set
      $ \widetilde{u} = \widetilde{v} \oplus
                         \widetilde{ (\quot{u}{v}) }  $.
\end{enumerate}

\begin{fakt}
  Let $u$ be a~representation of a~quantum group $(A,\Kom)$. Then
  the~representation $\widetilde{u}$ is uniquely determined up to
  an~equivalence.
\end{fakt}

\dow
  We first notice
  \begin{equation}
    \hlabel{q:vtrace}
    \tr \widetilde{u} = \tr u \:.
  \end{equation}

  Then let
  \begin{equation}
    \hlabel{q:Trozklad}
    \widetilde{u} \approx v_1 \oplus \ldots \oplus v_k
  \end{equation}
  be a~decomposition of $\widetilde{u}$ into irreducible components. Hence
  \[
    \tr u = \tr v_1 + \ldots + \tr v_k
  \]
  and ($\tr v_i$ are linearly independent if $v_i$ are nonequivalent
  irreducible representations) (\ref{q:Trozklad}) is uniquely determined.
\cnd

\vspace{2ex}
Let us introduce the~notation
\begin{equation}
   \hlabel{q:tptdef}
   \tpt = \: \WT \: \circ \tp \:.
\end{equation}

\begin{fakt}
  \hlabel{q:trUtrVfakt}
  Let $u$, $v$ be representations of a~quantum group, such that
  $\tr u \; \tr v = \tr v \; \tr u$ and $v \tp u$ is irreducible.
  Then $v \tp u \approx u \tp v$.
\end{fakt}

\dow
  One has
  \[
    \tr (u\tpt v) = \tr (u\tp v) =
    \tr u \; \tr v = \tr v \; \tr u =
    \tr (v\tp u) = \tr (v\tpt u) \:,
  \]
  which means that $u\tpt v \approx v\tpt u$.
  The~representation $v\tpt u \approx v\tp u$ is irreducible, therefore
  the~representation $u\tpt v$ is irreducible. Thus one has
  $u\tpt v \approx u\tp v$.
\cnd

One has $u \tpt v \approx \widetilde{u} \tpt \widetilde{v}$ for any two
representations $u$, $v$ (trace of both sides is the~same).


\section*{Acknowledgments}
\addcontentsline{toc}{section}{Acknowledgments}

We are very grateful to Prof.~S.~L.~Woronowicz for
setting the~problem and helpful discussions.


\addcontentsline{toc}{section}{References}

\begin{thebibliography}{9}
\bibitem{b:CL}
    De Concini, C., Lyubashenko, V.,
    Quantum function algebra at roots of~$1$,
    {\em Preprints di Matematica \/}
    {\bf 5},
    {\em Scuola Normale Superiore Pisa}, February 1993.
\bibitem{b:Lyakhovskaya}
    Gluschenkov, D.V., Lyakhovskaya, A.V.,
    Regualar Representation of \\ the~Quantum Heisenberg Double
    \{$\Uqsl \:, \mbox{Fun}_q(SL(2))$\} ($q$~is a~root of unity),
    {\em UUITP~---~27\slash 1993 \/},
    hep-th\slash 9311075.
\bibitem{b:Jimbo}
    Jimbo, M.,
    A~$q$--analogue of $U(gl(N+1))$, Hecke algebra,
    and the~Yang--Baxter equation.
    {\em Lett. Math. Phys. \/}
    {\bf 11}, 247-252 (1986).
\bibitem{b:Lusztig}
    Lusztig, G., Modular representations and quantum groups.
    {\em Contemporary Mathematics \/}
    {\bf 82}, 59-77 (1989).
\bibitem{b:CQGandTRR}
    Podle\'{s}, P.,
    Complex Quantum Groups and Their Real Representations.
    {\em Publ.\ RIMS, Kyoto University \/}
    {\bf 28}, 709-745 (1992).
\bibitem{b:RTF}
    Reshetikhin, N.Yu., Takhtadzhyan, L.A., Faddeev, L.D.,
    Quantization of Lie groups and Lie algebras.
    {\em Leningrad Math. J. \/}
    Vol.~1, No.~1, 193-225 (1990).
\bibitem{b:RA}
    Roche, P., Arnaudon, D.,
    Irreducible Representations of~the~Quantum Analogue of~$SU(2)$.
    {\em Lett. Math. Phys. \/}
    {\bf 17}, 295-300 (1989).
\bibitem{b:Parshall}
    Wang, J., Parshall, B.,
    Quantum linear groups.
    {\em Memoirs Amer. Math. Soc. \/}
    439, Providence, 1991.
\bibitem{b:SU2}
    Woronowicz, S.L.,
    Twisted $SU(2)$ group.
    An example of a~noncommutative differential calculus.
    {\em Publ.\ RIMS, Kyoto University \/}
    {\bf 23}, 117-181 (1987).
\bibitem{b:CompactMatrix}
    Woronowicz, S.L.,
    Compact Matrix Pseudogroups.
    {\em Commun. Math. Phys. \/}
    {\bf 111}, 613-665 (1987).
\bibitem{b:Differential}
    Woronowicz, S.L.,
    Differential Calculus on Compact Matrix Pseudogroups (Quantum Groups).
    {\em Commun. Math. Phys. \/}
    {\bf 122}, 125-170 (1989).
\bibitem{b:wykladWor}
    Woronowicz, S.L.,
    The lecture {\em Quantum groups} at Faculty of Physics,
    University of Warsaw (1990\slash 91)
\bibitem{b:SL2}
    Woronowicz, S.L.,
    New quantum deformation of $SL(2,\C)$. Hopf algebra level.
    {\em Rep. Math. Phys. \/}
    {\bf 30}, 259-269 (1991).
\bibitem{b:Lorentz}
    Woronowicz, S.L., Zakrzewski, S.,
    Quantum deformations of the~Lorentz group. The~Hopf $\ast$-algebra level.
    {\em Comp. Math. \/}
    {\bf 90}, 211-243 (1994).
\end{thebibliography}
\end{document}